\documentclass[preprint,12pt,3p]{elsarticle}

\newif\ifdraft
\drafttrue

\usepackage{amssymb}
\usepackage{amsmath}
\usepackage{soul}

\usepackage{booktabs}
\usepackage{subcaption}
\usepackage{url}
\usepackage{graphicx}
\usepackage{textcomp}
\usepackage{etex}
\usepackage{multirow}
\usepackage{float}
\usepackage{balance}
\usepackage{booktabs}
\usepackage{soul}
\usepackage{adjustbox}
\usepackage{subcaption}
\usepackage{comment}
\usepackage[inline]{enumitem}
\usepackage{makecell}
\usepackage{wrapfig}
\usepackage{lscape}

\newcommand{\subr}[1]{{\small\texttt{r/#1}}}

\usepackage{xcolor}
\usepackage{amsthm}%
\usepackage{mathrsfs}%
\usepackage{tabularx}
\usepackage{color}
\usepackage{colortbl}
\usepackage{rotating}
\usepackage{pifont}
\usepackage[acronym]{glossaries}
\usepackage{algorithm}
\usepackage{algpseudocode}
\usepackage{longtable}
\usepackage{tikz}
\usepackage[]{natbib}

\newcommand{\multicomment}[1]{}

\newcommand{\activity}{\texttt{activity}}
\newcommand{\diversity}{\texttt{diversity}}
\newcommand{\toxicity}{\texttt{toxicity}}

\newcommand{\ifdraftelse}[2]{%
 \ifdraft
 #1%
 \else
 #2%
 \fi
}

\DeclareMathOperator{\nmd}{NMD}
\DeclareMathOperator{\md}{MD}

\DeclareMathOperator*{\argmin}{arg\,min}

\definecolor{lightblue}{RGB}{220,230,241}
\definecolor{lightgreen}{RGB}{220,241,220}
\definecolor{lightpink}{RGB}{241,220,230}
\definecolor{lightyellow}{RGB}{255,250,205}
\definecolor{lightgray}{RGB}{235,235,235}
\definecolor{lightpurple}{RGB}{235,225,255}
\definecolor{lightorange}{RGB}{255,235,215}
\definecolor{lightteal}{RGB}{210,245,245}
\definecolor{lightred}{RGB}{255,220,220}

\definecolor{col-selected}{RGB}{178,223,138}
\definecolor{col-discarded}{RGB}{251,154,153}

\journal{Online Social Networks and Media}

\begin{document}

\begin{frontmatter}

%% Title, authors and addresses

%% use the tnoteref command within \title for footnotes;
%% use the tnotetext command for theassociated footnote;
%% use the fnref command within \author or \affiliation for footnotes;
%% use the fntext command for theassociated footnote;
%% use the corref command within \author for corresponding author footnotes;
%% use the cortext command for theassociated footnote;
%% use the ead command for the email address,
%% and the form \ead[url] for the home page:
%% \title{Title\tnoteref{label1}}
%% \tnotetext[label1]{}
%% \author{Name\corref{cor1}\fnref{label2}}
%% \ead{email address}
%% \ead[url]{home page}
%% \fntext[label2]{}
%% \cortext[cor1]{}
%% \affiliation{organization={},
%% addressline={},
%% city={},
%% postcode={},
%% state={},
%% country={}}
%% \fntext[label3]{}

% \title{Evaluating Feature Importance \\ 
% for Online Content Moderation \\ 
% via Quantification} 
\title{Quantifying Feature Importance \\ 
for Online Content Moderation} 

%% use optional labels to link authors explicitly to addresses:
%% \author[label1,label2]{}
%% \affiliation[label1]{organization={},
%% addressline={},
%% city={},
%% postcode={},
%% state={},
%% country={}}
%%
%% \affiliation[label2]{organization={},
%% addressline={},
%% city={},
%% postcode={},
%% state={},
%% country={}}
%\author[]{Anonymous Authors}

\def\thefootnote{*}\footnotetext{These authors contributed equally to this work.}
\author[unipi,iit]{Benedetta Tessa$^*$}
\author[isti]{Alejandro Moreo$^*$}
\author[iit]{Stefano Cresci}
\author[iit]{Tiziano Fagni}
\author[isti]{Fabrizio Sebastiani}

\affiliation[unipi]{organization={Dipartimento di Informatica, Università di Pisa},%Department and Organization
 addressline={Largo Bruno Pontecorvo 3},
 postcode={56127}, 
 city={Pisa},
 country={Italy}}
\affiliation[iit]{organization={Istituto di Informatica e Telematica, Consiglio Nazionale delle Ricerche},%Department and Organization
 addressline={ \\ Via Giuseppe Moruzzi 1}, 
 postcode={56124}, 
 city={Pisa},
 country={Italy}}
\affiliation[isti]{organization={Istituto di Scienza e Tecnologie dell'Informazione, Consiglio Nazionale delle Ricerche},%Department and Organization
 addressline={ \\ Via Giuseppe Moruzzi 1}, 
 postcode={56124}, 
 city={Pisa},
 country={Italy}}

%% Abstract
\begin{abstract}
Accurately estimating how users respond to moderation interventions is paramount for developing effective and user-centred moderation strategies. However, this requires a clear understanding of which user characteristics are associated with different behavioural responses, which is the goal of this work. We investigate the informativeness of 753 socio-behavioural, linguistic, relational, and psychological features, in predicting the behavioural changes of 16.8K users affected by a major moderation intervention on Reddit. To reach this goal, we frame the problem in terms of ``quantification'', a task well-suited to estimating shifts in aggregate user behaviour. We then apply a greedy feature selection strategy with the double goal of \textit{(i)} identifying the features that are most predictive of changes in user activity, toxicity, and participation diversity, and \textit{(ii)} estimating their importance. Our results allow identifying a small set of features that are consistently informative across all tasks, and determining that many others are either task-specific or of limited utility altogether. We also find that predictive performance varies according to the task, with changes in activity and toxicity being easier to estimate than changes in diversity. Overall, our results pave the way for the development of accurate systems that predict user reactions to moderation interventions. Furthermore, our findings highlight the complexity of post-moderation user behaviour, and indicate that effective moderation should be tailored not only to user traits but also to the specific objective of the intervention.
\end{abstract}

%%Graphical abstract
%\begin{graphicalabstract}
%\includegraphics{grabs}
%\end{graphicalabstract}

%%Research highlights
%\begin{highlights}
%\item Research highlight 1
%\item Research highlight 2
%\end{highlights}

%% Keywords

\end{frontmatter}

%% Add \usepackage{lineno} before \begin{document} and uncomment 
%% following line to enable line numbers
%% \linenumbers

%% main text
%%

%--------------------------------------------------------------------

%\tableofcontents

%--------------------------------------------------------------------

\section{Introduction}
\label{sec:intro}

The need for effective content moderation has become increasingly important for online platforms, in order to ensure the safety and well-being of online communities~\cite{gillespie2018custodians}. Without moderation, problematic phenomena such as the spread of mis- and disinformation~\cite{barnabo2023deep, ferrara2020misinformation}, political manipulation~\cite{cresci2014criticism, padinjaredathsuresh2024tracking}, toxicity~\cite{alvisi2025mapping, nogara2025longitudinal}, hate speech~\cite{giorgi2025human,pierri2024drivers}, and coordinated harmful behaviour~\cite{cima2024coordinated, mannocci2024detection}, risk threatening the integrity of the online discourse. In addition, platforms are legally required to implement moderation mechanisms in order to detect and mitigate certain types of harmful content~\cite{shahi2025year, tessa2025improving, trujillo2025dsa}. To moderate content and behaviours, platforms adopt a wide range of interventions that depend on the context and severity of the violation. These range from limiting the visibility of a post~\cite{chandrasekharan2022quarantined} to permanently removing a user from the platform~\cite{jhaver2021evaluating}. While these interventions are generally intended to restore the integrity of the online ecosystems, their effectiveness is not always guaranteed. Recent research has highlighted that some interventions may be ineffective, or even lead to unintended consequences, such as user churn, migration to less regulated platforms, or even an increase in toxicity~\cite{cima2025investigating, habib2022exploring, horta2021platform, trujillo2022make}.

One of the factors that currently limits the effectiveness of content moderation is the \textit{one-size-fits-all} approach to the deployment of moderation interventions, which do not account for individual user differences~\cite{cresci2022personalized}. Indeed, recent studies showed that different users respond in different ways to the same intervention, which reduces the overall effectiveness of user-independent moderation~\cite{trujillo2023one}. A crucial step towards user-centred or personalized moderation is the identification of those user characteristics that are responsible for the different reactions to moderation interventions. Therefore, in this work we aim to shed light on the main drivers behind behavioural change that follows a moderation intervention. A deeper understanding of what drives users to react in certain ways can pave the way to the development of personalized interventions, thus enhancing the overall efficiency and effectiveness of the content moderation process~\cite{cresci2022personalized}. To this end, we explore different behavioural and cognitive characteristics of a large set of social media users that were affected by a massive moderation intervention on Reddit. Among the features we consider are user activity, toxicity, writing style, emotional tone, and psychological traits. In order to assess which features are most relevant, we adopt an approach based on machine learning (ML). By training models and assessing the relative importance of different feature groups in predicting behavioural change after the intervention, we identify the most informative such groups.

Technically, our problem is complex due to the fact that the \textit{IID assumption},\footnote{The assumption that the training data and the test data are ``identically and independently distributed.''} on which essentially all supervised ML methods are based, is not verified here. This is because, in the context of online moderation, the set of data on which models are trained and the set of data to which the trained models are then applied are typically distributionally different~\cite{trujillo2022make, trujillo2023one}. %\tizfag{Qui diciamo che in questo contesto c'è un data shift nei dati ma non forniamo alcuna giustificazione sul perchè affermiamo questo. Potremmo accennare che abbiamo fatto una analisi preliminare ed è emerso questo problema, oppure citare qualche altro lavoro che menziona questa problematica specifica su conbtent moderation (ammesso che qualche lavoro di questo tipo esista) } \alemor{Secondo me si puó semplicemente assumere. Cioé, l'ipotesi di base è che qualcosa \emph{cambia} dopo un intervento. Perció quello che vediamo prima l'intervento e dopo è distribuito diversamente. Forse una frasetta a riguardo: ``The underlying assumption is that a moderation intervention produces behavioural changes in platform users, thereby effectively \emph{shifting} the distributions.''}
% The latter data are thus said to be \textit{out of distribution} (OOD) \alemor{ io non sono sicuro che OOD si applichi qua; poi ne parleremo}, and 
In these cases, the two distributions from which the two sets of data are sampled are said to be characterized by \textit{dataset shift}~\cite{Moreno-Torres:2012ay, Quinonero:2009kl, Storkey:2009lp}. The strategy we adopt thus consists of recasting our problem in terms of \textit{quantification}, the supervised learning task of predicting the prevalence of the classes in a set of unlabelled datapoints characterized by (some kind of) dataset shift~\cite{Forman:2008kx}. It is well known that solving this task by classifying all the unlabelled datapoints and checking how many have been assigned to each class (the ``classify and count'' method) is a na\"ive, suboptimal quantification method~\cite{Esuli:2023os}, since standard methods for training classifiers are based on the IID assumption. %, i.e., on the assumption that there is \textit{no} dataset shift.
Therefore, we adopt a strategy based on quantification methods because our problem can indeed be formulated in terms of class prevalence prediction, and because current quantification algorithms are, as hinted above, robust \textit{by construction} to dataset shift, or to some specific type of it. 
% \stecre{Briefly (1 sentence) remark why this approach is theoretically preferable to classify and count?} \fabseb{Fatto (see above).}
 
%--------------------------------------------------------------------

\paragraph{Contributions} To assess the main drivers of behavioural change, we implement and experiment with 753 features divided into 9 groups and 69 subgroups, reflecting a broad array of socio-behavioural, linguistic, relational, and psychological aspects of online human behaviour. We leverage these features to ``quantify'' the changes in activity, toxicity, and participation diversity that 16,828 users exhibited following the Great Ban, a massive deplatforming intervention occurred on Reddit in 2020~\cite{cima2025investigating}. Overall, our work provides the following main contributions:
\begin{itemize}
 \item We cast the prediction of changes in post-moderation behaviour as a quantification problem. Then, we combine it with a greedy feature selection algorithm to identify the most informative user-level features.
 \item We carry out an extensive empirical study involving 16.8K users and 753 features for three prediction tasks: quantify behavioral changes in activity, toxicity, and participation diversity following a moderation intervention. %\fabseb{Una cosa che non mi piace molto è che non adottiamo sempre una terminologia consistente; e.g., qui activity, toxicity, and diversity sono chiamate ``tasks'', e altrove (in modo più convincente) sono chiamate ``behavioural dimensions'' (``task'' lo usiamo anche per qualcos'altro). Mentre ciò potrebbe non generare confusione in chi è addentro nel mondo della content moderation, potrebbe generarla in un outsider. Analogmente, a volte parliamo di ``feature groups'', a volte di ``feature blocks'', etc. Io sarei per una radicale uniformazione della terminologia.}
 \item For each task, we identify the most informative set of features. Our results reveal a small set of features that are consistently informative, and shows that the rest are either task-specific or of limited utility.
 \item We discuss practical implications for effective moderation, including the need to adjust interventions to the context of the moderation, the aim of the intervention, and the characteristics of the moderated users.
\end{itemize}

%--------------------------------------------------------------------

%\subsection{Significance}
%
%\stecre{Decide if we want to have this section}

%--------------------------------------------------------------------

\section{Related Work}
\label{sec:relwork}
%--------------------------------------------------------------------

\subsection{Descriptive Content Moderation}
\label{sec:Descriptive}

Most research works on the effects of moderation interventions have adopted descriptive approaches (i.e., post-hoc observational designs) for estimating effects, focusing on how interventions influence user behaviour, community dynamics, and discourse visibility. Several works have examined community-level interventions such as bans, quarantines, and visibility restrictions. For example, analyses of ``The Great Ban'', a large-scale deplatforming intervention occurred on Reddit in 2020, revealed that subreddit removals produced heterogeneous effects across communities and user types, with top contributors often leaving and casual users changing the way they wrote~\cite{cima2024great, trujillo2021echo}. Further analyses showed that sequential interventions on \verb+r/The_Donald+ shaped user activity, toxicity, and information credibility, often inducing counter-intuitive effects such as increased polarisation~\cite{trujillo2022make}. Community-level interventions aimed at reducing the visibility of problematic user groups are the focus of~\cite{chandrasekharan2022quarantined, shen2022tale}, which demonstrate that such interventions decrease new user influx but do not significantly change the way existing users communicate. Finally, others analysed deplatforming interventions targeting influencers, and showed a significant drop in public attention towards them and a reduction of public presence of certain problematic ideas~\cite{horta2025deplatforming, jhaver2021evaluating}.

A complementary line of work focused on user-level, rather than community- or platform-level, behavioural responses to moderation. Cima et al.~\cite{cima2025investigating} identified substantial inter-user heterogeneity following a deplatforming intervention, with a small but notable group of users reacting negatively to the ban by posting highly toxic comments. 
Similar results are described in~\cite{trujillo2023one}, which found multiple outliers that diverged from community average estimates of intervention effects. Then, studies in~\cite{gleason2025suspense, yokotani2022} examined temporary suspensions, showing that punishment duration and user status influence recidivism and behavioural contagion among peers. Overall, these findings emphasize the need for moderation strategies tailored to user profiles.

%Studies on the deplatforming of influential users highlight how moderation actions can reshape the broader conversation and reduce the public presence of certain ideas. Jhaver et al.~\cite{jhaver2021evaluating} and Trujillo and Cresci~\cite{trujillo2022make} show that banning high-profile individuals reduces their ideological spread and overall toxicity in associated communities. Horta Ribeiro et al.~\cite{horta2025deplatforming} confirm that multi-platform deplatforming events lead to sustained decreases in public interest.

Lastly, some recent studies highlighted that user perceptions and bystander responses are increasingly recognized as critical in shaping moderation efficacy. Jhaver~\cite{jhaver2025bans} found that people are more likely to support bans and warnings if they believe harmful content affects others more than themselves, while Zhao and Hobbs~\cite{zhao2025effects} showed that moderation decisions voted by groups of users may help in reducing toxic comments and improving community standards. Other studies showed that user perceptions of moderation effectiveness vary between moderated and bystander users~\cite{cima2025contextualized, hong2024outcome}.

%--------------------------------------------------------------------

\subsection{Predictive Content Moderation}
\label{sec:Predictive}

Some works explored predictive approaches to content moderation, focusing not only on early identification of violations, but also on the consequences of moderation interventions, across different levels of analysis. One line of research focused on the preemptive detection of content violations. Kurdi et al.~\cite{kurdi2020video} analysed over 73,000 YouTube videos to identify features predictive of future removal, achieving promising results that support the feasibility of pre-emptive moderation. %Their machine-learned models achieve high accuracy (up to 92\%) in forecasting content removal, highlighting the feasibility of pre-emptive moderation tools. \fabseb{Io in genere sono contro al riportare cifre di accuratezza nelle introduzioni e nel related work, perché l'accuratezza che si riscontra dipende dal dataset, quindi le cifre che si riportano non sono informative.}
\cite{paudel2023lambretta} proposed \textsc{Lambretta}, a learning-to-rank system for the early identification of moderated tweets; by leveraging semantic keyword extraction from known misinformation tweets, \textsc{Lambretta} identified up to 20 times more policy-violating content than manual methods, with minimal human input. Finally, \textsc{Crossmod} is a cross-community moderation tool for Reddit that infers macro-norm violations via transfer learning from 100 subreddits, enabling effective detection in communities lacking historical moderation data~\cite{chandrasekharan2019crossmod}. At the community level,~\cite{habib2022proactive} proposed a framework to identify subreddits likely to face moderation restrictions. Their models use structural and behavioural indicators to predict future bans up to nine months in advance, thus enabling scalable proactive moderation.

Beyond content, other works aimed at predicting users and community effects of moderation interventions, ahead of their application. For example, this predictive approach was applied to detect and forecast user ban evasion on Wikipedia using behavioural and psycholinguistic features~\cite{niverthi2022characterizing}. Others tackled the task of forecasting user abandonment following a moderation intervention~\cite{tessa2025beyond}. The present work is particularly related to~\cite{niverthi2022characterizing} and~\cite{tessa2025beyond}, in that, similarly to the cited works, we carry out a ML task to predict the effects of a moderation intervention. However, contrary to previous work, we consider a broader range of possible reactions to an intervention, including changes in activity, toxicity, and diversity of user participation, and we do that with the aim of identifying the most informative features for predicting such effects, rather than for optimizing prediction accuracy. %While addressing the same prediction task, our work differs in that we significantly expand the set of features used and conduct an in-depth qualitative analysis, supported by quantification techniques, aimed at evaluating the effectiveness of distinct classes of features in predicting user abandonment.

Finally, some recent studies explored the utility of large language models (LLMs) for predictive moderation. Among them, Kumar et al.~\cite{Kumar_AbuHashem_Durumeric_2024} assessed the viability of using GPT-3.5 as a moderation agent. They applied it to 95 Reddit communities, finding that its performance varied significantly depending on community rules and content context. Others proposed the \textit{policy-as-prompt} paradigm, enabling moderation decisions directly from policy instructions embedded in LLM prompts~\cite{palla2025policy}. This approach reduces retraining needs but introduces challenges in prompt design and governance. Other works investigated small language models (SLMs) as lightweight alternatives to LLMs. The study in~\cite{zhan2024slm} demonstrated that fine-tuned SLMs may outperform LLMs in domain-specific moderation tasks; \cite{wang2025stand} introduced instead \textsc{STAND-Guard}, an SLM framework trained via cross-task learning that generalizes across novel moderation tasks.
%Jing at al.~\cite{jing2025text} integrate user-level embeddings with BERT encoders, showing that incorporating user behaviour and interaction patterns improves moderation accuracy. 

%--------------------------------------------------------------------

\subsection{Applications of Quantification}
\label{sec:Applications}

Quantification (a.k.a.\ ``learning to quantify'') is the supervised learning task of predicting the prevalence of the classes in a set of unlabelled datapoints characterized by dataset shift~\cite{Forman:2008kx}. Quantification finds applications in several downstream tasks whenever these latter need to be carried out under dataset shift. Examples of such tasks are improving the accuracy of classifiers~\cite{Saerens:2002uq}, measuring the fairness of classifiers~\cite{Fabris:2023pj} and rankers~\cite{Jaenich:2025rr}, predicting the accuracy of classifiers~\cite{Volpi:2025bx, Volpi:2025zm}, and calibrating classifiers~\cite{Moreo:2025vn}. 

Aside from these tasks, quantification is useful in all the disciplines that have an inherent focus on \textit{aggregate} data (i.e., that are focused on phenomena at the ``macro'' level) and where the object of study are \textit{populations} and not individuals, as in our case. In these fields, the prevalence estimates are interesting in themselves, and not just for solving a different task. Examples of such fields are epidemiology~\cite{King:2008fk}, the social sciences and political science~\cite{Hopkins:2010fk}, market research~\cite{Sebastiani:2019xg}, and ecological modelling~\cite{Gonzalez:2019fh}. While no applications of quantification to online content monitoring have yet been discussed in the literature, the application we deal with in this paper squarely falls in this category. % (to the best of our knowledge)

Many quantification methods have been proposed to date (\cite{Esuli:2023os,Gonzalez:2017it} are two recent surveys on the subject). Most such methods have been tested under (and proved their value for addressing) prior probability shift, often considered the ``paradigmatic case'' of dataset shift~\cite{Ziegler:2024qq}. Instead, other types of dataset shift, such as covariate shift and concept shift, have received considerably less attention in the quantification literature~\cite{Gonzalez:2024cs, Tasche:2022hh}.

%--------------------------------------------------------------------

\section{Quantifying Feature Importance for Online Content Moderation}
\label{sec:Methodology}

%--------------------------------------------------------------------

\subsection{Problem Definition}
\label{sec:problem}

Our overarching goal is to identify the set of features most relevant in determining user reactions to a moderation intervention, and to estimate their degree of relevance. To reach this goal, we develop a quantification model capable of estimating the class distribution that represents behavioural variations in an online community following an intervention. In particular, we consider changes across multiple dimensions of user behaviour: activity, toxicity, and the diversity of user participation across subreddits. Quantifying changes in each of these dimensions constitutes a distinct task, requiring separate models trained to capture the specific dynamics associated with each behavioural aspect. In the remainder of the paper, we refer to each behavioural dimension as a distinct task. 
This multi-dimensional approach enables a more comprehensive assessment of feature relevance, as it allows us to identify whether certain feature groups are informative in general, or only in relation to specific types of behavioural change. The adoption of a quantification-based approach allows us to assess the informativeness of features at the population level, focusing on how well they capture aggregate shifts in behaviour, rather than at the level of individual predictions. This approach is well suited to the moderation context, where the objective is to understand broad user trends and effects, rather than to classify individual responses with high precision. %Although our final goal is to develop a quantification model capable of estimating the class distribution that represents behavioural variations in an online community after a moderation intervention, in this work we focus on a preliminary step towards this goal, i.e., identifying \emph{which features} give rise to a more effective quantification system.
%\fabseb{Qui specificherei che identificare queste features non serve solo ad addestrare il quantification model di cui sopra, ma serve a identificare quali sono in genere le features più utili per la content moderation; lo facciamo usando la quantification perché la bontà di una feature si vede al livello aggregato, e non al livello individuale.}

%--------------------------------------------------------------------

\subsubsection{Class Definition}
\label{sec:classdefinition}

%For each behavioural dimension (i.e., activity, toxicity, diversity), the classes in our tasks correspond to discrete levels of variation in that dimension.
For each task (i.e., activity, toxicity, diversity), the classes correspond to discrete levels of variation in that dimension.
In detail, we define the following five-point scale: \textsf{HighlyDecreased}, \textsf{ModeratelyDecreased}, \textsf{NoVariation}, \textsf{ModeratelyIncreased}, \textsf{HighlyIncreased}.\footnote{The way in which these classes are computed is thoroughly described in Section~\ref{sec:labelling} ``Ground-truth Labelling.''}  This 5-point scale generates an \textit{ordinal} quantification context, one in which a total order is defined on our set of  classes. As a consequence, mis-assigning some prevalence mass---say, from \textsf{HighlyDecreased} to \textsf{HighlyIncreased}---is a more serious mistake than mis-assigning the same prevalence mass from \textsf{HighlyDecreased} to \textsf{ModeratelyDecreased}. %``very low variation'', ``low variation'', ``no variation'', ``high variation'', and ``very high variation.''
%We consider several independent problems, namely, the variation in activity, toxicity, and diversity levels of users in the social network (more on this in Section~\ref{sec:effects}).

The three behavioural dimensions that define the tasks are measured as follows. 
%We consider several independent problems, namely, the variation in activity, toxicity, and diversity.
Given a fixed period of time, \textit{Activity} is measured as the total number of comments posted by a user in that time period; \textit{Toxicity} is defined as the 75th percentile of toxicity scores across the user’s comments posted during that time period, and \textit{diversity} measures the variety of communities (i.e., subreddits) in which a user has actively participated during that time period. The latter is computed via the well-known \textit{Hill diversity index}~\cite{trujillo2023one}
\[
{}^{q}H = \left( \sum_{i=1}^{R} p_i^q \right)^{\frac{1}{1 - q}}
\]
where \( p_i \) is the proportion of comments in subreddit \( i \), and \( R \) is the total number of distinct subreddits the user participated in, with parameter $q=1.5$.

%--------------------------------------------------------------------

\subsubsection{Task Formulation}
\label{sec:task_formulation}

We frame the problem as a supervised learning task. Accordingly, we assume access to a dataset $D = \{(x_i, y_i)\}_{i=1}^N$, where $x_i \in \mathcal{X} \subset \mathbb{R}^M$ represents a user in the social network. Here, $x_i$ is an $M$-dimensional vector of covariates, while $y_i \in \mathcal{Y}$ represents the (true) label assigned to $x_i$, indicating one of the five variation levels for a particular behavioural aspect. For brevity, we encode these five levels as $\mathcal{Y} = \{1, 2, 3, 4, 5\}$. A vector $x$ of covariates is a concatenation of different feature blocks $x = (f_1; \ldots; f_F)$, with $f_i \in \mathbb{R}^{d_i}$ a subvector of $d_i$ real-valued components, such that $\sum_{i=1}^F d_i = M$. In our case we have $M=753$ and $F=69$.

The main objective is thus to select a subset of feature blocks that results in a more effective quantification system. Formally, if we consider $I = \{1, \ldots, F\}$ the set of indices of all available feature blocks, our goal can be stated as identifying
\begin{equation}
S^* = \argmin_{S \subset I} \mathcal{L}(q(S), D)
\end{equation}
where $\mathcal{L}$ is the loss function measuring the performance of a quantifier $q$ applied to instances from $D$ restricted to the selected features $x' = (f_{s_1}; \ldots; f_{s_{|S^*|}})$, with $s_i\in S^*$. Since the number of possible feature block combinations is $2^{69} \approx 6 \times 10^{20}$, the problem  of examining all possible combinations of feature blocks is obviously intractable. To explore this space we thus resort to search heuristics, as detailed in Section~\ref{sec:greedy}.

%--------------------------------------------------------------------

%\subsubsection{Loss Function: Quantification Accuracy as a Proxy of Feature Importance}
\subsubsection{Quantification Accuracy as a Proxy of Feature Importance}
\label{sec:proxy}
We treat the accuracy  of our quantification model as a proxy of the quality of the set of feature blocks on which it was trained. This accuracy is measured in terms of a loss function, which must thus reflect the performance of the quantifier in a broad scenario. 

To this end, we split the dataset $D$ into a labelled set $L$ (consisting of 8,000 randomly drawn instances), to be used to train the quantification model, and an unlabelled set $U$ (the remaining instances), to be used for evaluation purposes. In order to reduce the experimental bias caused by one specific partition, we repeat this process 5 times. %\fabseb{Non mi piace che mettiamo numeri prima della sezione sperimentale.}

To reflect scenarios characterised by different training set sizes, we divide the training set $L$ into 16 batches $L_1, \ldots, L_{16}$ of 500 instances each. This setup allows us to monitor the behaviour of the quantifier in a series of experiments where we use, as the training data (here noted as $T$), progressively higher portions of $L$, from 500 ($T_1=L_1$), 1,000 ($T_2=L_1 \cup L_2$), $\ldots$, up to 8,000 instances ($T_{16}=L=\cup_{i=1}^{16} L_i$). % For each training size, a performance metric is computed, and the area under the resulting learning curve (across the 16 points) is used as a global performance measure.
%\stecre{Qui si inizia a parlare del modo in cui valutiamo le performance, ma il discorso rimane un po' appeso fino a Sez~\ref{sec:expsetting} dove si introduce il metodo di evaluation. Visto che Sez~\ref{sec:expsetting} e' breve, valutiamo se non convenga metterla direttamente qui, almeno le informazioni presentate in questo punto sarebbero complete, e potrebbe risultare tutto piu' chiaro per chi legge.} \alemor{Ho spostato qua la parte di Evaluation Measure, come suggerivi.}

Since our ultimate goal is to obtain a quantifier that performs well in estimating variations in prevalence, we subject each experiment to a \emph{stress test}, in which we evaluate the quantifier according to its ability to accurately predict widely varying class prevalence values. To this aim, we adopt the \emph{Artificial Prevalence Protocol} (APP)~\cite{Esuli:2023os, Forman:2008kx}, in which the test set $U$ is used to generate many test samples with different class distributions. Specifically, we generate 1,000 test samples $U_1,\ldots,U_{1000}$ as follows. We first generate 1,000 random prevalence vectors $\mathbf{p}_1, ..., \mathbf{p}_{1000}$, uniformly sampled from the probability simplex $\Delta^4$, corresponding to the space of all possible distributions over 5 classes (which has 4 degrees of freedom) using the Kraemer sampling algorithm~\cite{smith2004sampling}. For each vector $\mathbf{p}_i$, we draw from $U$ a test sample $U_i$ of 500 instances that satisfies the specified prevalence values. The quantifier is then asked to estimate the class distribution of the 1,000 test samples, thus producing a series of estimates $\hat{\mathbf{p}}_1, ..., \hat{\mathbf{p}}_{1000}$. As our evaluation measure, we use the mean quantification error across all 1,000 test samples. %This protocol is a standard procedure to evaluate quantification models under varying prevalence conditions.

%--------------------------------------------------------------------

\subsection{Evaluation Measures}
\label{sec:expsetting} 
%\fabseb{Bisogna controllare se la notazione del resto del papero è consistente con la notazione di questa sezione.} \fabseb{Io questa parte sulle misure di valutazione la metetrei in una sezione a parte e la sposterei nella sezione sperimentale.}
Given that we tackle a set of ordinal quantification tasks, we measure the prediction error the models incur into when estimating the true prevalence $\mathbf{p}$ as $\hat{\mathbf{p}}$, by means of a standard evaluation measure for ordinal quantification---namely, via the \emph{Normalized Match Distance} (NMD), defined in~\cite{Sakai:2018cf} as
\begin{align}
 \begin{split}
 \label{eq:NMD}
 \nmd(\mathbf{p},\hat{\mathbf{p}}) = &
 \frac{1}{n-1}\md(\mathbf{p},\hat{\mathbf{p}})
 \end{split}
\end{align}
where $\frac{1}{n-1}$ is a normalization factor that allows NMD to range between 0 (best prediction) and 1 (worst prediction). Here, $n$ is the number of available classes (in our case, $n=5$), and MD is the well-known \emph{Match Distance}~\citep{Werman:1985tm}, defined as
\begin{align}
 \begin{split}
 \label{eq:EMD}
 \md(\mathbf{p},\hat{\mathbf{p}})
 = & \sum_{i=1}^{n-1} d(y_{i},y_{i+1})\cdot
 |P(y_{i})-\hat{P}(y_{i})|
 \end{split}
\end{align}
where $\smash{P(y_{i})=\sum_{j=1}^{i}p(y_{j})}$ is the
prevalence of $y_{i}$ in the cumulative distribution of $\mathbf{p}$,
$\hat{P}(y_{i})= \sum_{j=1}^{i}\hat{p}(y_{j})$ is an estimate of it,
and $d(y_{i},y_{i+1})$ is the ``semantic distance'' between
consecutive classes $y_{i}$ and $y_{i+1}$, i.e., the cost we incur in
mistaking $y_{i}$ for $y_{i+1}$ or vice versa. Throughout this paper,
we assume $d(y_{i},y_{i+1}) = 1$ for all $i\in\{1, 2, 3, 4\}$. 
MD is \textit{the} standard measure of evaluation in ordinal quantification%OQ
~\citep{Bunse:2024uo, Castano:2024ug}. %, where it is often called \emph{Earth Mover's Distance} (EMD). In fact, MD is a special case of EMD as originally defined in~\cite{Rubner:1998kx}.
%%% FEATURE IMPORTANCE %%
Since, as mentioned above, for each choice of a group $S$ of features and a task $T$ we run 16 experiments (one for each training set size in \{500, 1000, ..., 7500, 8000\}), we compute the error incurred by the system when using feature group $S$ for task $T$ as the mean of the NMD scores across these 16 setups, which we here denote by $\operatorname{MNMD}_{S}^{T}(\mathbf{p},\hat{\mathbf{p}})$. In order to evaluate the overall contribution of feature group $S$ for task $T$, we perform ablation experiments on task $T$ in which we remove group $S$ from the set of all features, and measure this contribution by the \textit{relative increase in error} (RIE) that derives from removing feature group $S$ from the set of all features

\[
\operatorname{RIE}_{S'}^T = \frac{\operatorname{MNMD}_{S/S'}^T - \operatorname{MNMD}_{S}^T}{\operatorname{MNMD}_{S}^T}
\]

%\fabseb{Sì, ma qui sopra non diciamo neanche di quale curva parliamo; io rimpiazzerei questa parte (da ``Finally'' in poi) con la sottostante.}

%Since, as mentioned above, for each choice of a group $S$ of features and behavioural dimension $B$  we run 16 experiments (one for each training set size in \{500, 1000, ..., 7500, 8000\}), we compute the error incurred by the system when using feature group $S$ for $B$ as the mean of the NMD scores across these 16 setups, which we here denote by $\operatorname{MNMD}_{S}^{B}$. In order to evaluate the overall contribution of feature subgroup $S'\in S$ for the behavioural dimension $B$ with respect to a reference group $S$, we perform ablation experiments on $B$ in which we remove group $S'$ from $S$, and measure this contribution by the \textit{relative increase in error} (here denote by RIE), i.e.
%

%--------------------------------------------------------------------

\subsubsection{Greedy Exploration for Feature Block Selection}
\label{sec:greedy}

The greedy exploration for selecting features consists of carrying out a series of tests in which the contribution of each feature block is evaluated. The exploration begins from an initial \emph{candidate} configuration (i.e., an initial selection of feature groups) $C\subseteq I$. The initial configuration is selected by picking the best-performing configuration across different setups $G_1,\ldots,G_9$, each consisting of a specific superset of the 9 groups of feature groups that we will describe in Section~\ref{sec:featureextraction}, plus one additional configuration $G_0=I$ that contains all indexes. % (e.g., all the Writing-style features, or all the Relational features, etc.)
Note that the groups $G_1,\ldots,G_9$ form a partition over the feature groups, that is $G_i \cap G_j=\emptyset$ if $i\neq j$ and $\bigcup_{i=1}^9G_i=G_0=I$.

The exploration then goes on by evaluating different feature group in turns. Each evaluation of a feature block $f_i$ consists of testing whether removing the index $i$ from $C$ (if $i\in C$) or whether adding the index $i$ to $C$ (if $i\notin C$) reduces the loss. In such case, the modification is retained, and otherwise is reverted. Since the number of feature groups is relatively high, and it is likely that unimportant groups are discarded early in the process, we restrict feature addition to the first three rounds of exploration. The subsequent rounds focus on \emph{refining} the selected set of features (presumably reduced) and only test for feature removal. Refinement continues until no further improvement is observed.

The order in which the feature groups are presented corresponds to the increasing performance (i.e., descending loss) of a system trained using only that feature block individually. In other words, the order of $f_i$ corresponds to the inverse rank of $\mathcal{L}(q(\{i\}),D)$ across all $i\in I$. This means that the least promising candidates, those yielding the highest loss, are processed first, while the most promising ones come last. The entire process is repeated through three rounds, and the final configuration $C$ is returned as an estimate $\hat{S}$ of $S^*$. The pseudocode describing this greedy exploration is presented in Algorithm~\ref{alg:greedy}. 
%\hl{spiegare meglio algoritmo}

\begin{algorithm}[t]
\caption{Greedy selection of feature groups.}\label{alg:greedy}
\begin{algorithmic}
\footnotesize
\State $C\gets\argmin_{G_i \in \{G_0,\ldots,G_9\}} \mathcal{L}(q(G_i), D)$ \Comment{Initial configuration}
\State $\text{best-loss}=\mathcal{L}(C,D)$
\State $\text{sorted}\gets\text{sort}(\{f_1,\ldots,f_{69}\}, \mathcal{L}, D)$ \Comment{Sort candidate feature groups}

\State $\text{improvement}\gets$ True
\State $\text{round}\gets 0$
\While{improvement}
 \State $\text{improvement}\gets$ False
 \For{$f'_i \in \text{sorted}$}
 \If{$f'_i\in C$} 
 \State $C' \gets C - f'_i$ \Comment{Remove the feature block}
 \Else
 \If{$\text{round}<3$}
 \State $C' \gets C \cup f'_i$ \Comment{Add the feature block (only during the first 3 rounds)}
 \Else
 \State \textbf{continue} \Comment{The refinement rounds (round $\geq$ 3) only consider ablation tests}
 \EndIf 
 \EndIf
 \State $\text{new-loss}\gets\mathcal{L}(C', D)$
 \If{$\text{new-loss}<\text{best-loss}$}
 \State $C \gets C'$ \Comment{Keep the last change} 
 \State $\text{best-loss} \gets\text{new-loss}$
 \State $\text{improvement}\gets$ True
 \EndIf
 \EndFor
 \State $\text{round}\gets \text{round}+1$
\EndWhile \\
\Return $C$
\end{algorithmic}
\end{algorithm}
 
%--------------------------------------------------------------------

\subsection{Feature Extraction}
\label{sec:featureextraction}
We extracted a total of 753 features, divided into the following categories: \texttt{EMBEDDINGS} (384, 51.0\%), \texttt{LIWC} (112, 14.9\%), \texttt{WRITING\_STYLE} (81, 10.8\%), \texttt{RELATIONAL} (36, 4.8\%), \texttt{ACTIVITY} (34, 4.5\%), \texttt{SENTIMENT} (32, 4.2\%), \texttt{SOC\_PSY} (31, 4.1\%), \texttt{TOXICITY} (24, 3.2\%), and \texttt{EMOTIONS} (19, 2.5\%). Within each group, features sharing similar characteristics were further divided into subgroups. This hierarchical organisation allows for a fine-grained analysis of which specific subdimensions are most predictive. Each category of features is briefly introduced in the following. Additionally, a detailed summary of all groups of features and their corresponding subgroups is provided in Table \ref{tab:features_split} in the Appendix.

%--------------------------------------------------------------------

\subsubsection{Embeddings}

This group of features is the most numerous, as it contains more than 50\% of the overall total. For each user, we computed a 384-dimensional embedding vector for each comment using the sentence transformer model \textsc{all-MiniLM-L6-v2}\footnote{\url{https://huggingface.co/sentence-transformers/all-MiniLM-L6-v2}}, and we then averaged these vectors to obtain a single representation per user.
Embeddings are widely used due to their ability to capture semantic and linguistic information. In particular, they have proven particularly effective in tasks related to detecting harmful content on social media, such as hate speech detection~\cite{ santhiya2024comparative}, deepfake identification~\cite{fagni2021tweepfake}, and cyberbullying~\cite{kumar2023cyberbullying}.
Since these tasks aim to understand user behaviour on social media, embeddings could also be effective in capturing linguistic patterns that are indicative of how users react to an intervention, hence our experimentation.

%--------------------------------------------------------------------

\subsubsection{Linguistic Inquiry and Word Count}

Linguistic Inquiry and Word Count (LIWC) is a lexicon that maps textual content to psycholinguistic categories, including cognitive processes, social concerns, and affective expression~\cite{boyd2022development}. LIWC has been widely used in the literature due to its ability to capture psychologically meaningful aspects of language. For instance, it has been effectively used in tasks such as hate speech detection~\cite{raut2024enhancing}, misinformation detection~\cite{gambini2024anatomy}, and personal value prediction~\cite{silva2021predicting}. For this reason, LIWC can offer valuable insights into the psychological dimensions that may influence how users react to moderation interventions. We use the latest version of the lexicon, i.e., LIWC-22.\footnote{\url{https://www.liwc.app/}} 

%--------------------------------------------------------------------

\subsubsection{Writing Style}

Previous studies on the effects of content moderation suggest that users may modify their linguistic style in response to moderation interventions. For instance, users who continued to engage on the platform were found to shift from first- to third-person plural pronouns~\cite{horta2021platform}, while other research reported a decrease in ingroup-oriented language~\cite{trujillo2021echo}.
Inspired by these insights, this group of features includes part-of-speech distributions and readability indices, which may help detect subtle linguistic changes following moderation.

%--------------------------------------------------------------------

\subsubsection{Activity}
\label{sec:feat:activity}

A fundamental aspect of user behaviour is their activity on social media, as platforms' revenues strongly depend on user engagement~\cite{carlson2020you, habib2022exploring}. However, certain interventions caused users to reduce their activity or even abandon the platform~\cite{cima2025investigating, cima2024great, trujillo2023one, trujillo2021echo}, which justifies the adoption of this group of features. In fact, previous work already showed that activity trends, number of comments, as well as temporal indicators, are among the most predictive features of user churn following moderation interventions~\cite{tessa2025beyond}.

%--------------------------------------------------------------------

\subsubsection{Relational}

At the core of any social media platform lies the interaction between users and communities, as these are the drivers of user activity and engagement. However, moderation interventions may change these interactions~\cite{habib2022proactive, trujillo2022make}. Following this intuition, we include a group of relational features aimed at capturing how users interact with communities. For instance, we measure a user’s influence in both banned and non-banned communities as well as initiative and adaptability indicators~\cite{mazza2022investigating}. The rationale behind these features stems from the idea that a user’s level of involvement in non-banned communities, as well as their interactions with other users affected by the ban, may influence their response to a moderation intervention.

%--------------------------------------------------------------------

\subsubsection{Sentiment}

Sentiment analysis is a widely adopted task in social media research, as it allows to assess the positive, negative, or neutral tone and sentiment of user-generated content.
In particular, it is widely used in hate speech detection, as negative polarity often correlates with toxicity~\cite{anjum2024hate}.
Here, we computed user sentiment with VADER~\cite{hutto2014vader}, a rule-based sentiment analyser specifically developed for social media. It assigns to each comment a sentiment score that can either be negative, neutral, positive, and a compound score that summarizes overall sentiment. To obtain user-level features, we aggregated these scores across all comments posted by each user. Similarly, we computed the sentiment of the emojis used by each user, leveraging the dictionary developed in~\citep{kralj2015sentiment}, which returns the same scores returned by VADER for each emoji.% \fabseb{``returns the same scores for each emoji''? Che vuol dire? Immagino che nel dizionario a ogni emoji sia associato uno score?}

%--------------------------------------------------------------------

\subsubsection{Socio-Psychological}

Predicting the effects of moderation interventions is related to predicting user behaviour, which is known to be influenced by psychological and demographic factors~\cite{drazkiewicz2022study, williams2017individual}. Additionally, several studies have shown that such characteristics are associated with online behaviours, including toxic and aggressive tendencies~\cite{cresci2022personalized, giorgi2025human}. On top of that, recent studies have demonstrated that incorporating socio-psychological features leads to significant improvements in hate speech detection tasks~\cite{raut2024enhancing}.
Among the features included in this group are users’ demographic attributes, such as age and gender, inferred by training a classifier on the \textsc{Pandora} dataset~\cite{vukojevic2021pandora}. We also computed personality traits based on the OCEAN model (Openness, Conscientiousness, Extraversion, Agreeableness, Neuroticism), obtained by averaging the scores returned for each comment using a pre-trained classifier.\footnote{\url{https://github.com/jkwieser/personality-prediction-from-text}} Finally, for each user we assessed their moral values as resulting from their comments, by using the \textsc{Moral Foundations} framework, a well-established theory that identifies five core dimensions of moral reasoning: care/harm, fairness/cheating, loyalty/betrayal, authority/subversion, and purity/degradation~\cite{kwak2021frameaxis}. For each comment, we extracted for each dimension two scores, i.e., bias, which indicates the direction (e.g., toward care or harm), and intensity, expressing how strongly a dimension is conveyed. We then averaged each score for each user.

%--------------------------------------------------------------------

\subsubsection{Toxicity}
\label{sec:feat:toxicity}

One of the most common reasons behind the ban of online communities is the spread of toxic content~\cite{habib2022proactive,horta2025deplatforming,trujillo2021echo}. Hence, measuring user toxicity levels prior to the ban may serve as a strong predictor of their post-ban behaviour. Here, we computed toxicity scores with \textsc{Detoxify}~\cite{Detoxify}, a widely adopted multilingual deep learning classifier, largely used in tasks related to the study or detection of toxicity on social media~\cite{cima2024great, gilda2022predicting, tessa2025beyond}. For each comment, \textsc{Detoxify} returns a range of scores such as toxicity, severe toxicity, obscenity, insult, identity attack, and threat that we aggregated at the user level. 

%--------------------------------------------------------------------

\subsubsection{Emotions}

This group includes features aimed at capturing the emotions expressed by users in their comments, which can be indicative of their psychological state and may shape their response to moderation. Emotion analysis have been successfully used in tasks such as hate speech detection or assessing the wellbeing of online users~\cite{raut2024enhancing}. Here, we leveraged two lexicon-based resources: the \textsc{NRC Emotion Intensity Lexicon} (NRC-EIL)~\cite{LREC18-AIL}, which provides word-level intensity scores for anger, anticipation, disgust, fear, joy, sadness, surprise, and trust, and the \textsc{NRC Valence-Arousal-Dominance} (NRC-VAD)~\cite{vad-acl2018} lexicon, which quantifies words along continuous affective dimensions of valence (positivity), arousal (intensity), and dominance (control). 
 To complement these lexicon-based features, we incorporated \textsc{EmoAtlas}~\cite{semeraro2025emoatlas}, a computational framework that assigns a \textit{z}-score to each comment for discrete emotions including joy, trust, fear, surprise, sadness, disgust, anger, and anticipation.

%--------------------------------------------------------------------

%--------------------------------------------------------------------

\section{Experimental Settings}
\label{sec:experiments}

Here we describe the experiments we have carried out. The code that reproduces our experiments is available on GitHub.\footnote{\url{https://github.com/AlexMoreo/EffectPrediction}} % \stecre{In questa sezione abbiamo 3 sottosezioni. 4.1 che introduce la metrica di valutazione, e 4.2 e 4.3 che sono relative al dataset. Propongo di spostare 4.1 dentro sezione 3 e qui di lasciare solo le sottosezioni relative al dataset. Credo che renderebbe il paper meglio organizzato e più omogeneo.} \alemor{Concordo; l'ho spostata a sezione 3.}

%--------------------------------------------------------------------

%--------------------------------------------------------------------

\subsection{Dataset}
\label{sec:dataset}

\begin{table}[t]
\small
\centering
\setlength{\tabcolsep}{3pt}
\begin{tabular}{ l r r r }
\toprule
\textbf{subreddit} & \textbf{subscribers} & \textbf{users} & \textbf{comments}\\
\midrule
\subr{ChapoTrapHouse} & 159,185 & 9,295 & 1,368,874 \\
\subr{The\_Donald} & 792,050 & 4,262 & 619,434 \\
\subr{ConsumeProduct} & 64,937 & 1,730 & 60,073 \\
\subr{DarkHumorAndMemes} & 421,506 & 1,632 & 35,561 \\ 
\subr{GenderCritical} & 64,772 & 1,091 & 94,735 \\
\subr{TheNewRight} & 41,230 & 729 & 5,792 \\
\subr{soyboys} & 17,578 & 596 & 5,102 \\
\subr{ShitNeoconsSay} & 8,701 & 559 & 9,178 \\
\subr{DebateAltRight} & 7,381 & 488 & 27,814 \\
\subr{DarkJokeCentral} & 185,399 &316 &  3,214\\
\subr{Wojak} &  26,816 & 244 & 1,666\\ 
\subr{HateCrimeHoaxes} &  20,111 & 189 & 775\\
\subr{CCJ2} & 11,834 & 150 & 9,785\\
\subr{imgoingtohellforthis2} & 47,363  & 93 &376 \\
\subr{OandAExclusiveForum} & 2,389 & 60 & 1,313 \\ 
\bottomrule
\end{tabular}
\caption{List of banned subreddits used in this study. For each subreddit, we indicate the number of subscribers, the number of unique users, and the total number of comments.} 
\label{tab:dataset}
\end{table}

For our analyses we leverage the dataset described in~\cite{cima2025investigating,cima2024great}. It consists of approximately 16M Reddit comments posted by 16,828 users affected by the Great Ban on June 29, 2020. The dataset spans a 14 months period of time, including seven months before and seven months after the ban, which enables thorough studies of the ban's effects. %The authors then filtered out those users
The users in the dataset represent those who were consistently active prior to the ban in at least one of 15 large banned subreddits. %These 15 subreddits include the ten largest ones publicly disclosed by Reddit at the moment of the ban, and five additional ones identified in~\cite{trujillo2021echo}, who filtered out private communities and those with fewer than 2,000 active users.
Table \ref{tab:dataset} lists the 15 selected subreddits along with the number of subscribers and active users therein. The dataset does not include bots, who were filtered out via a semi-automatic procedure based on filtering rules and manual validation~\cite{cima2024great}. The 16M user comments include 2.2M comments that the 16.8K users posted in the seven months prior to the ban \textit{within} the banned subreddits, and 13.8M comments that the same users posted in the 14 months period \textit{outside} of the banned subreddits. Since the selected subreddits were permanently shut down with the ban, no post-ban activity exists within them. Therefore, user activity outside of the banned subreddits is necessary to assess the effects of the intervention.

Based on this dataset, we computed user features using comments posted prior to the ban both inside and outside of the banned subreddits. Instead, the labelling of users (i.e., as those who increased/decreased activity/toxicity/diversity) was based solely on their comments outside the banned subreddits, for which we have both pre- and post-ban data.
%The dataset includes all comments posted in the selected subreddits from December 2019 to June 2020, totalling 8 million comments by around 194,000 users. The dataset was then filtered to retain only users who posted at least once per month, and bots were removed through automatic rules and manual validation. This process resulted in a final set of 2.2 million comments by 16,828 users. In addition, the authors collected comments made by these users outside the banned subreddits over a 7-month period before and after the ban. Finally, to ensure robustness in predicting the effects in terms of toxicity and diversity, we only considered users who remained relatively active after the ban and posted more than 10 comments during both the pre-ban and post-ban periods.
Additionally, to avoid possible spurious fluctuations, when computing features we excluded all comments made on the day of the ban. As a result, 753 features were computed for 16,522 users. Finally, since both toxicity and diversity are computed from user comments, their reliability depends on having a sufficient level of post-ban activity~\cite{cima2025investigating}. Therefore, to ensure the robustness of our measurements, we restricted the labelling of toxicity and diversity changes to users who posted at least ten comments in the post-ban period.\footnote{Activity in the pre-ban period is guaranteed by design, since the original dataset only contains users who were consistently active before the ban~\cite{cima2024great}.}
%
%\multicomment{
\begin{figure}[t]
 \centering
 \begin{subfigure}[t]{0.475\textwidth}
 \centering
 \includegraphics[width=\linewidth]{ 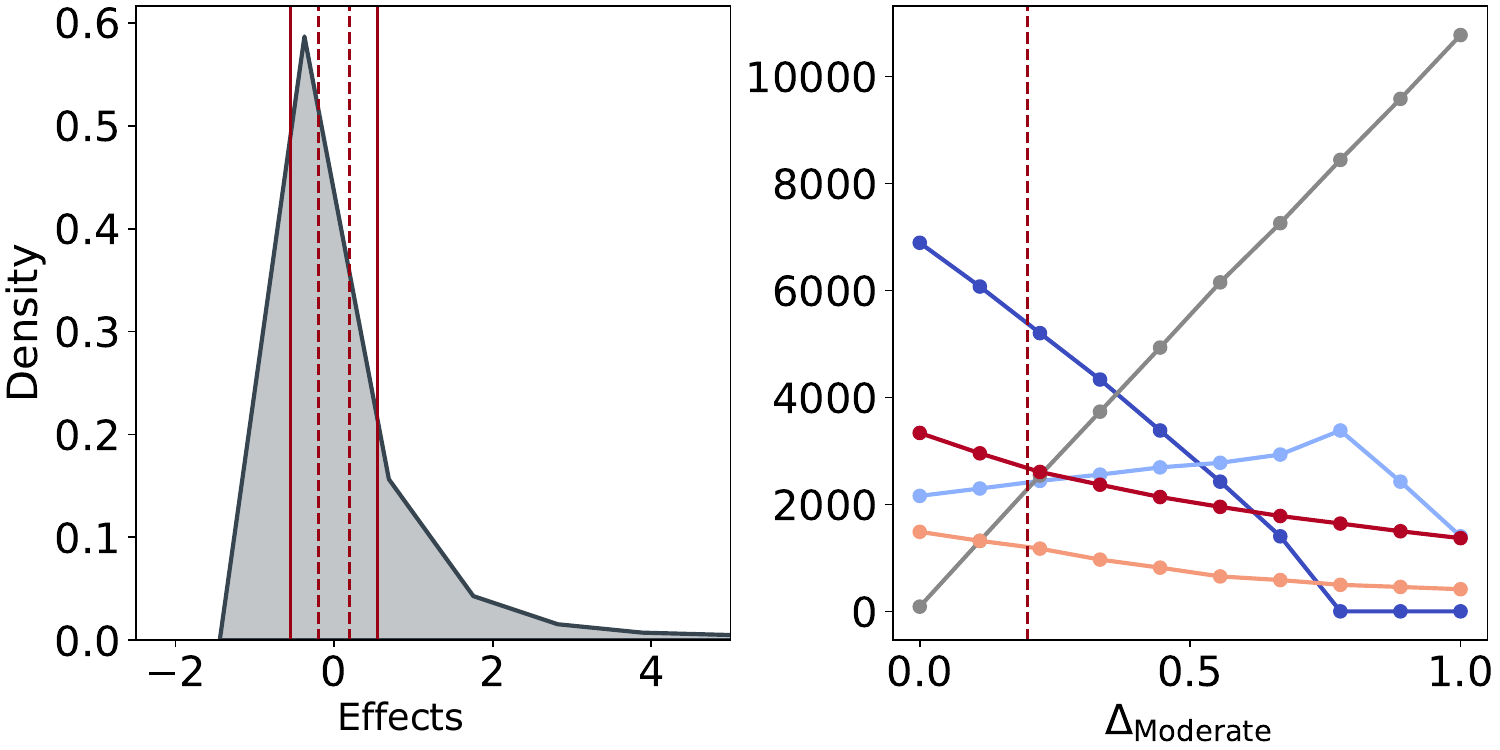}
 \caption{\activity}
 \end{subfigure}
 \hfill
 \begin{subfigure}[t]{0.475\textwidth}
 \centering
 \includegraphics[width=\linewidth]{ 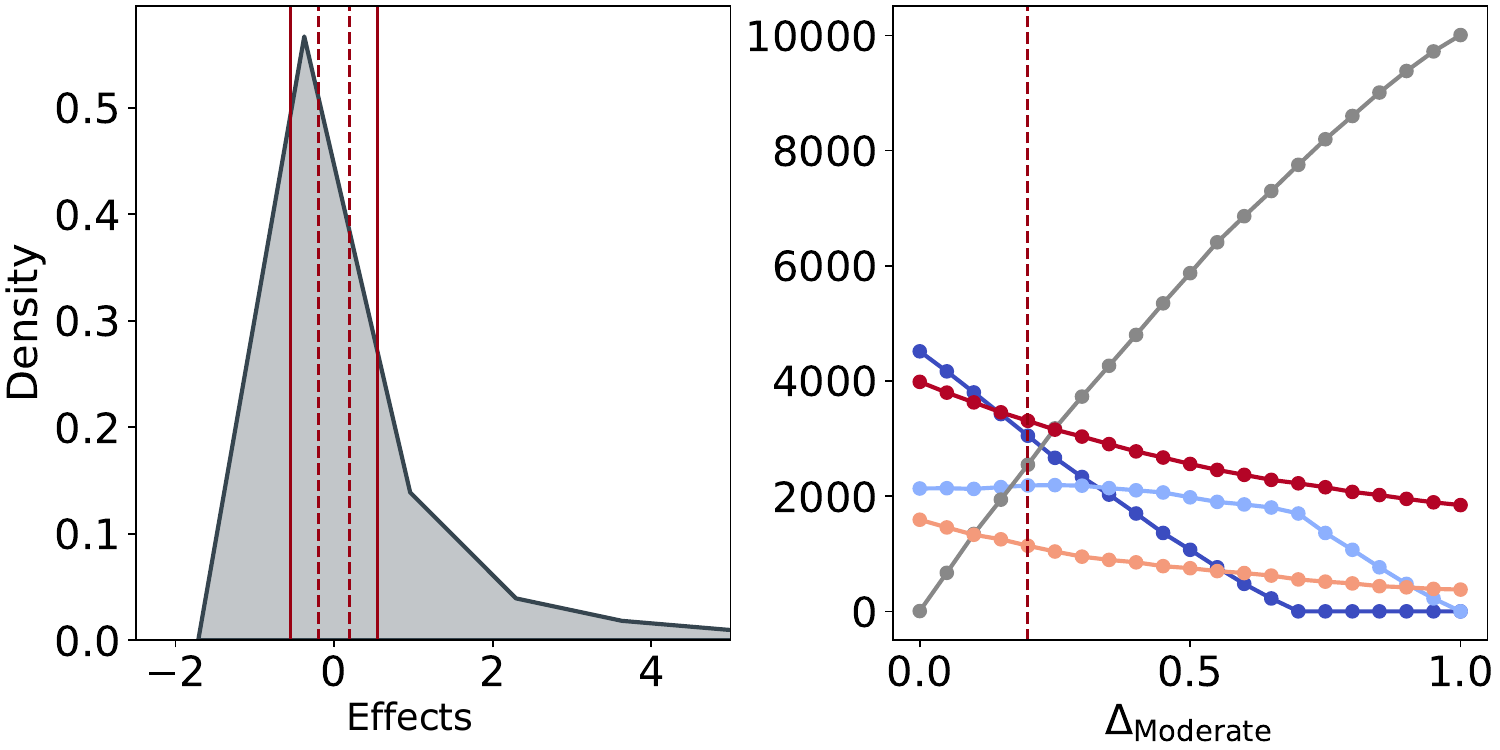}
 \caption{\toxicity}
 \end{subfigure}
 %\vspace{0.5em}
 \begin{subfigure}[t]{0.675\textwidth}
 \centering
 \includegraphics[width=\linewidth]{ 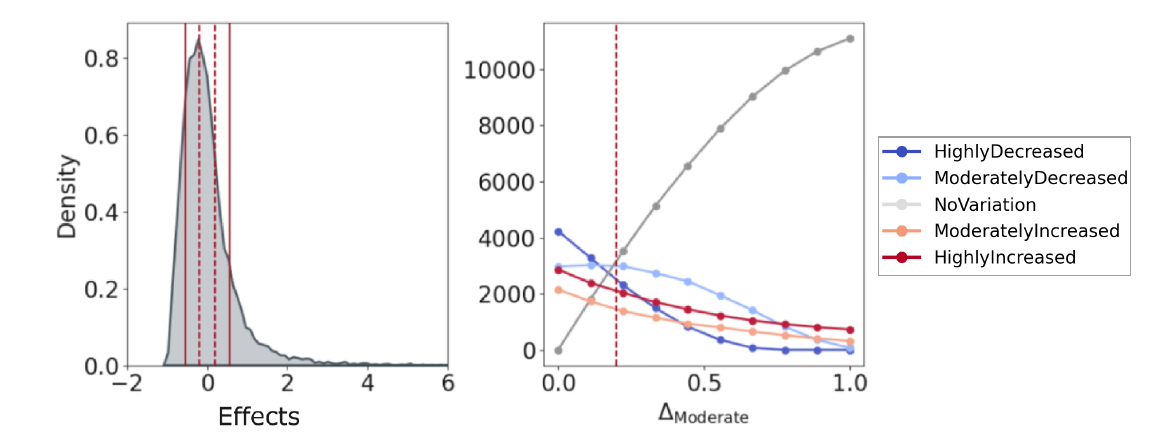}
 \caption{\diversity}
 \end{subfigure}
 \caption{Effects distribution and class distribution as a function of the parameter $\Delta$ for each of the considered tasks: activity, diversity, and toxicity. For the effect scores, we report the kernel density estimation (KDE) of the continuous values. The red dashed lines indicate the thresholds $\pm \Delta_{\text{Moderate}}$, while the red solid lines indicate $\pm \Delta_{\text{High}}$. %\fabseb{Bisogna cambiare la legenda della figura, in modo da usare i nomi corretti delle classi.}
 }
 \label{fig:distributions}
\end{figure}
%}

%--------------------------------------------------------------------

\subsection{Ground Truth Labelling}
\label{sec:labelling}

%To evaluate the effects of the intervention, for each user we compute the relative change they exhibit along three behavioural dimensions, according to the measure
%\[
%\text{Effect}_B = \frac{B_{\text{post}} - B_{\text{pre}}}{B_{\text{pre}}} \quad \text{with } B \in \{\text{activity}, \text{toxicity}, \text{diversity}\}
%\]

To evaluate the effects of the intervention, for each user we compute the relative change they exhibit along three tasks, according to the measure
\[
\text{Effect}_T = \frac{T_{\text{post}} - T_{\text{pre}}}{T_{\text{pre}}} \quad \text{with } T \in \{\text{activity}, \text{toxicity}, \text{diversity}\}
\]
The resulting values are continuous numbers, which we need to discretize into five classes in order to carry out the supervised ordinal quantification task. To this end, we define two thresholds $\Delta_\text{Moderate}$, $\Delta_\text{High} > 0$, with $\Delta_\text{Moderate} < \Delta_\text{High}$. Then, for each task $T$, we label users as
\[
\text{Label}_T = 
\left\{
\begin{array}{lcrcl}
 \text{\textsf{HighlyIncreased}} & \text{if} & \Delta_\text{High} < & \text{Effect}_T & \\
 \text{\textsf{ModeratelyIncreased}} & \text{if} & \Delta_\text{Moderate} < & \text{Effect}_T & \le \Delta_\text{High} \\
 \text{\textsf{NoVariation}} & \text{if} & -\Delta_\text{Moderate} \le & \text{Effect}_T & \le \Delta_\text{Moderate} \\
 \text{\textsf{ModeratelyDecreased}} & \text{if} & -\Delta_\text{High} \le & \text{Effect}_T & < -\Delta_\text{Moderate} \\
 \text{\textsf{HighlyDecreased}} & \text{if} & & \text{Effect}_T & < -\Delta_\text{High}
\end{array}
\right.
\]
The distributions of user-level effects $\text{Effect}_T$ across the three %behavioural dimensions 
tasks approximately follow Gaussian curves centred around zero, indicating that most users experienced only modest changes, while fewer exhibited pronounced increases or decreases. When defining the $\Delta_\text{Moderate}$ and $\Delta_\text{High}$ thresholds for class labels, we considered two key criteria. First, thresholds were chosen to reflect practically meaningful changes in user behaviour, in line with the real-world implications of user churn, rising toxicity, or shifting participation patterns. Second, we aimed to produce a relatively balanced class distribution, ensuring that all five classes were sufficiently represented. %This not only enhances the interpretability of the results but also makes the ordinal quantification task more robust and informative.
Figure~\ref{fig:distributions} shows the class distributions, in terms of number of users per class, for each %behavioural dimension
task, as a function of $\Delta_\text{Moderate}$, with $\Delta_\text{High} = \Delta_\text{Moderate} + 0.35$. Based on this preliminary analysis, we set $\Delta_\text{Moderate} = 0.2$ and $\Delta_\text{High} = 0.55$ for all tasks. %behavioural dimensions. 
%Therefore, we define two thresholds: $\Delta$ and $\Theta$ with $\Theta$ = $\Delta$ + 0.35. Specifically, any relative change below -$\Theta$ is treated Highly Decreased; changes between - $\Theta$ and -$\Delta$ indicate a Moderate Decrease, those falling between -$\Delta$ and $\Delta$ indicate No variation, values from $\Delta$ to $\Theta$ suggest a Moderate Increase and values exceeding $\Theta$ correspond to a High Increase. Choosing a value of $\Delta$ too far from zero would result in very few users falling into the class with no significant change in $D$. On the other hand, a value too close to zero would make moderate or strong changes indistinguishable. Moreover, we aimed to achieve a balanced class distribution. Following this rationale and the distributions shown in Figure \ref{fig:distributions}, we set $\Delta = 0.20$ and $\Theta = 0.55$ for all three behavioural dimensions considered.

%--------------------------------------------------------------------

\section{Results}
\label{sec:Results}
%\hl{This section presents the results of our analyses, starting from those obtained with the full set of features up to those related to our feature selection.}
%In this section, we turn to discuss the experimental results we have obtained. In Section~\ref{sec:quantifierchoice} we discuss the choice of the quantifier. In Section~\ref{sec:initialisation} we turn to describe the initial configuration of the greedy optimisation, while in Section~\ref{sec:ordering} we report on the individual evaluation of feature blocks that gives rise to the ordering of inspection. Finally, in Section~\ref{sec:finalselection} we report the resulting selection of feature blocks for each dataset.

%--------------------------------------------------------------------

%\subsubsection{Choosing a Representative Quantification Method}
\subsection{Quantifying with the Full Feature Set}
%\hl{Aggiungi come abbiamo valutato l'importance delle features}
\label{sec:quantifierchoice}

We begin by evaluating the performance of several well-known quantification methods using the full set of features available. This allows us to compare how effectively each method estimates class distributions across the three tasks---prediction of changes in activity, toxicity, and participation diversity---when provided with maximum information. The best-performing method from this comparison is then adopted for all subsequent experiments on feature selection. We follow this strategy because repeating the entire feature evaluation pipeline for multiple quantifiers would come at a considerable computational cost. The results presented here thus serve both to identify an overall strong-performing quantifier and to establish a point of reference against which improvements from feature optimization can be measured.

We consider the following three well-known quantification methods:
\begin{itemize}
 \item \texttt{CC}: Classify and Count is a weak quantifier that simply \textit{(i)} trains a classifier on the training data, \textit{(ii)} uses the classifier to issue label predictions for all the instances in the test sample, and \textit{(iii)} computes the percentages of test items attributed to each class. 
 \item \texttt{PACC}: Probabilistic Classify and Count~\citep{Bella:2010kx} is a quantifier that corrects an initial raw count estimate by taking into account the class-conditional error rates of the classifier (estimated via cross-validation). In contrast to \texttt{CC}, the underlying classifier of \texttt{PACC} is a probabilistic one that returns estimates of posterior probabilities for each data point.
 \item \texttt{EMQ}: Expectation Maximisation for Quantification~\cite{Saerens:2002uq} %(sometimes called SLD after the name of its proponents)\alemor{ se decidiamo di chiamarlo SLD dovrei rigenerare le figure di tabella 2, ma è semplice},
 is a mutually recursive iterative algorithm that alternates between two steps: \textit{(i)} the \emph{expectation} step, in which the class prevalence values are estimated as the average of the posterior probabilities, and \textit{(ii)} the \emph{maximisation} step, in which the posterior probabilities are updated to reflect the most recent class prior estimates. This process is repeated until convergence.
\end{itemize}
In addition to these, we also considered more recent methods such as \texttt{KDEy}~\citep{Moreo:2025lq}, but ultimately excluded them due to their high computational complexity. Since our overarching goal is not to achieve the best possible predictive performance but to assess feature importance, we prioritize methods that offer a good trade-off between effectiveness and scalability. %First, we carry out an initial experiment in which we select, out of a set of candidate quantification systems, the most promising one. To this aim, we consider three well-known quantification methods as potential candidates:\footnote{During preliminary experiments, we also tested more recent quantification methods, including KDEy~\citep{Moreo:2025lq}. However, due to its high training cost, running it across our full set of experiments would have been computationally prohibitive.}

All three above quantifiers rely on an underlying classification model to generate label predictions or probabilities, which are then used to estimate class prevalence. Following a common trend in the quantification literature, we rely on Logistic Regression (\texttt{LR}) for this purpose. We optimize the hyperparameters of \texttt{LR} independently for each quantifier and training set, via grid-search exploration, following the quantification-oriented model selection protocol proposed in~\cite{Moreo:2021sp}. %\footnote{This model selection protocol is designed to optimize hyperparameters for the task of quantification, rather than classification. To this end, the classifier’s hyperparameters are treated as part of the quantifier, and are therefore evaluated based on their ability to accurately estimate class prevalence, rather than to predict individual class labels.} In our experiments, we explore the regularisation hyperparameter $\mathcal{C}$ of LR in the range $\{10^i\}_{i=-4}^{i=4}$, and whether to rebalance (by setting the ``ClassWeight'' hyperparameter to ``Balanced'') the datapoints in order to counter the possible class imbalance of the training sample. For the implementation of LR we rely on \texttt{scikit-learn}~\citep{Pedregosa:2011yo}, while for the implementations of CC, PACC, and EMQ we rely on \texttt{QuaPy}~\citep{Moreo:2021bs}.
We rely on \texttt{scikit-learn} for the implementation of \texttt{LR} and on \texttt{QuaPy}~\citep{Moreo:2021bs} for the implementations of \texttt{CC}, \texttt{PACC}, and \texttt{EMQ}.

In addition to the above quantifiers trained with the full feature set, we also report the results of the Maximum Likelihood Prevalence Estimator (\texttt{MLPE}), a naïve method that assumes training and test data are IID. \texttt{MLPE} simply takes the class prevalence observed in the training data as an estimate for any incoming test set, regardless of its actual distribution. Finally, we also report the results of \texttt{EMQO}, an optimized \texttt{EMQ} model equipped with the best feature groups identified for each dataset with the feature selection process. Although the feature selection process is discussed in a later section, here we include the results of \texttt{EMQO} as a point of reference, serving as a lower bound on quantification error and providing context for interpreting the performance of models that use the full feature set. 
%This figure also reports the results of the \emph{Maximum Likelihood Prevalence Estimator} (MLPE), a naïve method that assumes training and test data are IID. As recalled in Section~\ref{sec:proxy}, our experiments simulate prior shift using APP. This way, 

\begin{figure}[t]
 \centering%
 \includegraphics[width=.33\textwidth]{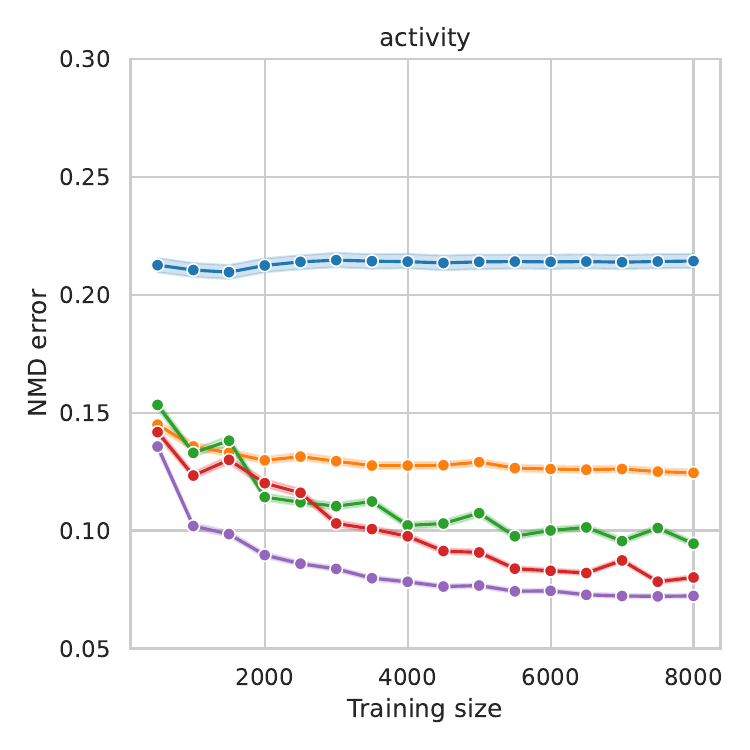}%
 \includegraphics[width=.33\textwidth]{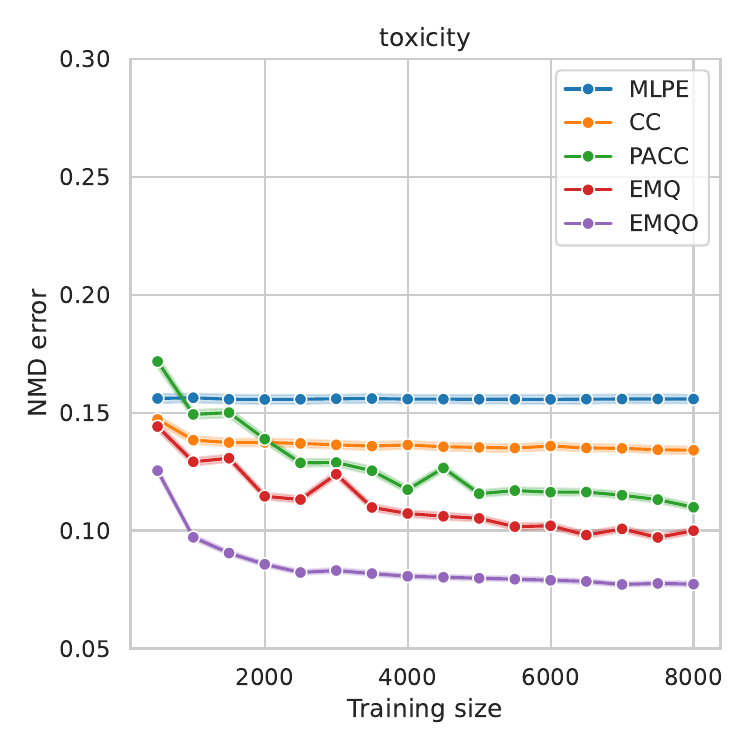}%
 \includegraphics[width=.33\textwidth]{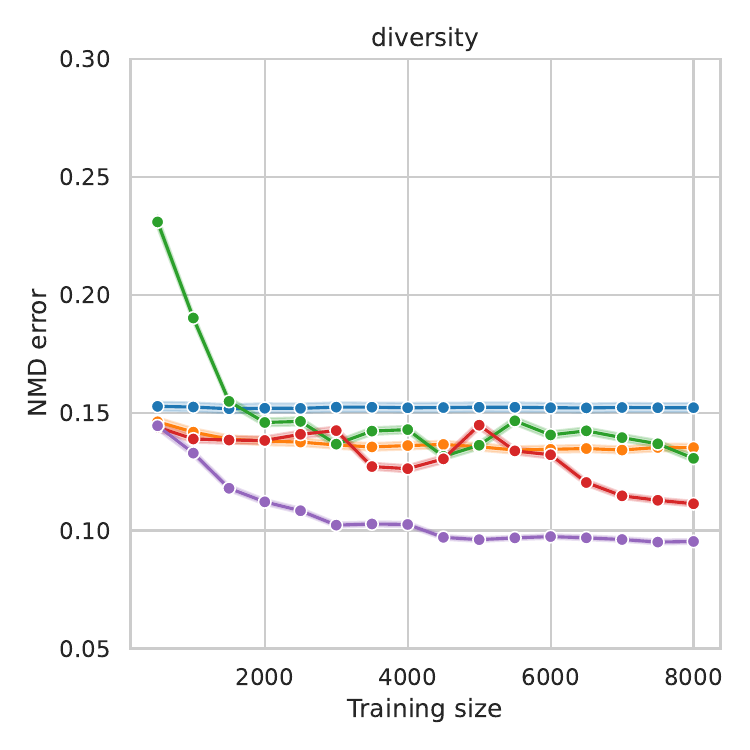}%
 \caption{Quantification error of the models as a function of training set size. The curves report the error trends across incremental subsets for the \activity, \toxicity, and \diversity\ tasks.} %\stecre{Consiglio di ingrandire leggermente il font nelle figure in modo che si leggano meglio. Forse conviene ingrandire leggermente anche le linee ed i punti. Inoltre lascerei una sola legenda (nella prima sottofigura, o in quella dove c'è più spazio), tanto sono tutte uguali. Nel testo ho rinominato EMQ \textit{optimized} in EMQO (for consistency): se vi piace andrebbe aggiornata l'etichetta nella legenda.} \alemor{Ho fatto la modifica EMQO, tolto 2 legende, e messe in verticale; secondo me cosi il font si vede già bene, ma volendo lo si puó ingrandire ulteriormente.}
 \label{fig:curves}
\end{figure}

Figure~\ref{fig:curves} displays the quantification error of each model as the training set size increases, allowing us to assess how performance scales with data availability. Each curve shows the trend in error across incremental training subsets, from 500 to 8,000 instances, for all three behavioural tasks. The figure reveals that \texttt{MLPE} and \texttt{CC} yield poor estimates of class prevalence. Instead, \texttt{PACC} and, especially, \texttt{EMQ}, attain much lower errors. In addition, while the simpler models exhibit a relatively stable performance independently of the training set size, the most powerful ones consistently reduce the quantification error when provided with more training data. These results are robust to the task at hand. Based on these results, in the following sections we adopt \texttt{EMQ} as our quantifier of choice for estimating feature importance. In addition to the previous findings, Figure~\ref{fig:curves} also shows a marked and consistent improvement of \texttt{EMQO} with respect to \texttt{EMQ} in all three tasks, which demonstrates the goodness of the feature selection process and the usefulness of an optimized set of features.
%This plot also shows the learning curve of EMQ \emph{optimized}; these lines correspond to the performance trend of EMQ when equipped with the best feature blocks we have identified for each dataset. While the selection process will be described in the sections to come, these plots already show a large improvement with respect to EMQ, i.e., with respect to a system that uses all feature blocks.
\begin{figure}[t]
 \centering
 \includegraphics[width=\textwidth]{ 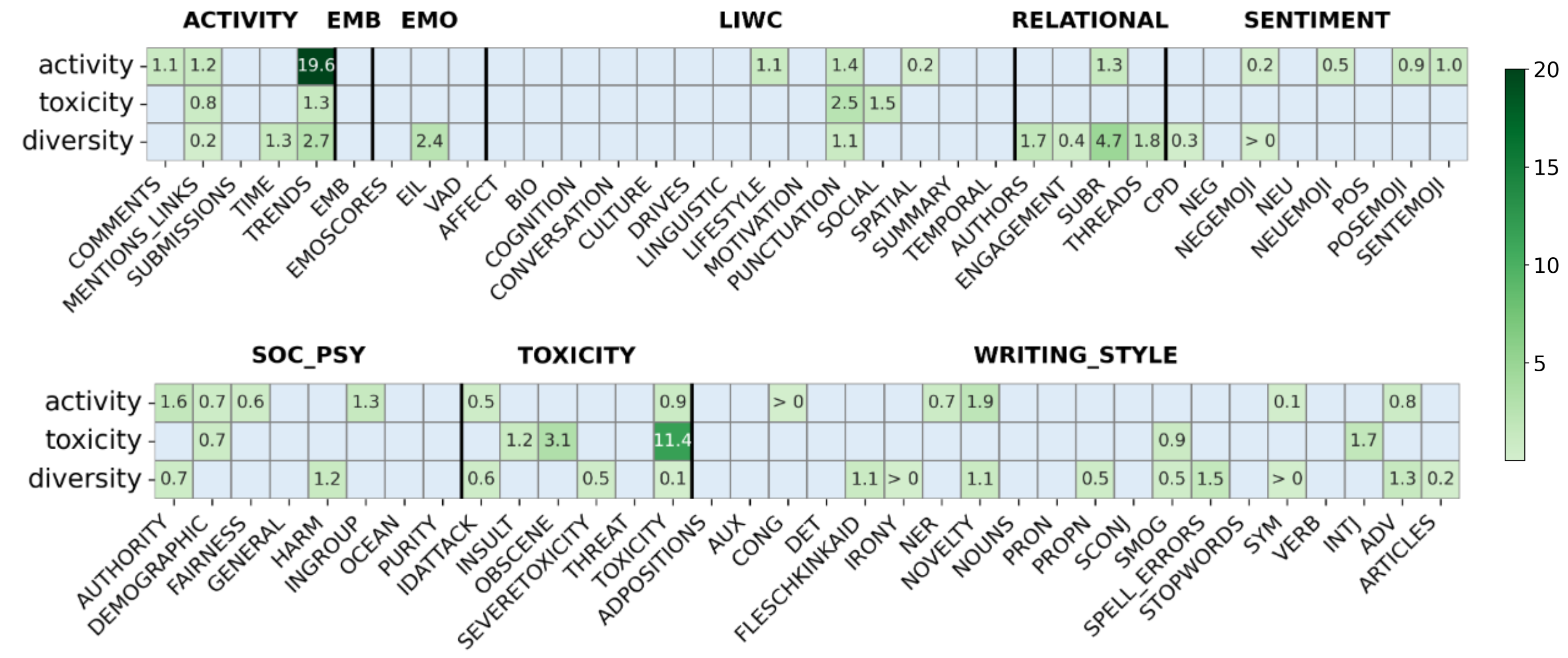}
 \caption{Percentage $\text{RIE}_{S}^T$ for each subgroup across each task. Darker green-coloured cells indicate higher importance of a subgroup for a given task, whereas blue-coloured cells denote that the subgroup was not used in that task.}
 \label{fig:feat-imp}
\end{figure}

%--------------------------------------------------------------------

\subsection{Selecting the Most Informative Features}
\label{sec:feature-selection}
\begin{table}[t]
\small
\centering
    \begin{tabular}{lrrrrr}
        \toprule
        & \multicolumn{3}{c}{\textbf{task}} & \multicolumn{2}{c}{\textbf{overall}} \\
        \cmidrule{2-4} \cmidrule(ll){5-6}
        \textbf{feature group} & \multicolumn{1}{c}{\textit{activity}} & \multicolumn{1}{c}{\textit{toxicity}} & \multicolumn{1}{c}{\textit{diversity}} & \textit{mean} & \textit{stdev} \\
        \midrule
        \texttt{ACTIVITY} & \textbf{711.4} & 1198.5 & 1171.7 & 1027.2 & 273.8 \\
        \texttt{EMBEDDINGS} & 1025.4 & 948.1 & \underline{986.3} & \underline{986.6} & 38.7 \\
        \texttt{EMOTIONS} & 1376.9 & 1230.6 & 1211.8 & 1273.1 & 90.4 \\
        \texttt{LIWC} & 1077.3 & 987.9 & 1058.3 & 1041.2 & 47.1 \\
        \texttt{RELATIONAL} & 795.3 & 1188.0 & 995.3 & 992.9 & 196.4 \\
        \texttt{SENTIMENT} & 1182.3 & 914.8 & 1292.6 & 1129.9 & 194.3\\
        \texttt{SOC\_PSY} & 1166.2 & 1060.8 & 1174.1 & 1133.7 & 63.3 \\
        \texttt{TOXICITY} & 947.8 & \textbf{733.1} & 1311.7 & 997.5 & 292.5 \\
        \texttt{WRITING\_STYLE} & 977.6 & 1004.9 & 1019.3 & 1000.6 & 21.2 \\
        \midrule
        \texttt{ALL} & \underline{748.8} & \underline{830.5} & \textbf{984.1} & \textbf{854.5} & 119.5 \\
    \bottomrule
    \end{tabular}
\caption{Group-wise initial exploration. Bold indicates the best result for each dataset, in terms of MNMD, and also indicates the initial feature block used as a reference value for the heuristic feature selection procedure. 
%Red cells indicate the worst performance and green cells the best; intermediate values are shown using a color gradient interpolated between these two extremes.
}
\label{tab:initial_choice}
\end{table}

%\begin{table}[t]
% \centering
% \resizebox{1\textwidth}{!}{
% \input{tables/root}
% }
% \caption{Group-wise initial exploration. Bold indicates the best result for each dataset, in terms of AUC, and also indicates the initial feature block used as a reference value for the heuristic feature selection procedure. Red cells indicate the worst performance and green cells the best; intermediate values are shown using a colour gradient interpolated between these two extremes.}
% \label{tab:initial_choice}
%\end{table}
\paragraph{Initializing the Feature Selection Algorithm} Before running the greedy feature selection algorithm, we must choose an initial configuration of feature groups from which the search begins. This is done by evaluating the performance of each feature group individually, along with the full feature set, and selecting the configuration that yields the lowest quantification error. The initial evaluation is carried out by running the evaluation procedure described in Section~\ref{sec:proxy}.
This provides a strong starting point for the iterative search algorithm that follows. % As discussed in Section~\ref{sec:greedy}, our greedy exploration departs from an initial configuration, chosen among all the configurations corresponding to feature groups (see Section~\ref{sec:featureextraction}), plus one configuration accounting for all features. 

Table~\ref{tab:initial_choice} reports the quantification error made by each configuration in each task. %\stecre{@AM: Esattamente come è calcolato questo quantification error? E' la AUC dell'errore? Visto che la usiamo in vari punti, non vale la pena definirla in Sezione~\ref{sec:expsetting}? Vale la pena chiamarla tipo eAUC (error AUC) per distinguerla dalla più comune AUC-ROC? La distinzione è importante perché la nostra è migliore se bassa, al contrario dell'altra. Ho paura che così possa creare confusione.} \alemor{Anche qua, ogni esperimento vol dire eseguire tutto il processo descritto in Section~\ref{sec:proxy}. Per AUC hai ragione: possiamo dire Area Under the Error Curve (AUEC) oppure Integrated Error (IE). Fabrizio ne sa di più di misure di evaluation, facciamolo scegliere a lui.} 
In table, the best result (i.e., lowest error) for each task and for the mean performance across all tasks is shown in bold, and the second-best is underlined. As shown, the best configuration for the activity and toxicity tasks correspond to the namesake feature groups. Instead, for the task of predicting changes in participation diversity, the best configuration is the one consisting of \texttt{ALL} feature groups. The results in Table~\ref{tab:initial_choice} highlight some interesting patterns. As expected, the feature groups most directly related to the target variables, \texttt{ACTIVITY} and \texttt{TOXICITY}, are the best predictors for their respective tasks, confirming the relevance of those features. In contrast, no single group is clearly aligned with the diversity task, and as a result, using \texttt{ALL} feature groups together yields the best performance in that case. Interestingly, the full feature set also ranks second-best for both activity and toxicity, suggesting that while other groups contain useful information, combining everything indiscriminately is suboptimal, which underscores the importance of feature selection. Moreover, the overall results in table show that certain feature groups, like \texttt{EMBEDDINGS}, perform consistently well across tasks, while others are highly specialized. Examples of the former are the aforementioned \texttt{ACTIVITY} and \texttt{TOXICITY} features, which exhibit large variability in performance. Finally, the consistently higher quantification errors observed for the diversity task indicate that it is inherently more difficult given the available features.
%ACTIVITY (Section~\ref{sec:feat:activity}) and TOXICITY (Section~\ref{sec:feat:toxicity}), respectively.
%This is sensible, given that no feature group is specifically designed to represent diversity-related aspects\alemor{check questa affermazione}. Additionally, this table also shows that ``diversity'' appears to be the most challenging dataset, as evidenced by the fact that the loss achieved by the best configuration is significantly higher than in the other two cases.

%--------------------------------------------------------------------

\paragraph{Ordering the Subgroups of Features} Once the initial set of features for each task has been selected as per the results of Table~\ref{tab:initial_choice}, the 69 subgroups of features that make up our full feature set are processed sequentially in order to assess if and how much they contribute to the overall prediction, separately for each task. The order in which the subgroups of features are processed corresponds to their performance when used in isolation. %\stecre{Qual è il razionale per usare questo ordinamento? E' possibile verificare se c'è una correlazione tra l'ordine con cui vengono processati i sottogruppi e il fatto che vengano scelti?} \alemor{L'idea era che si partiva da ``ALL'', e quindi era meglio levare subito i peggiori. Quando ho pensato a scegliere uno ``starting point'' migliore, in realtá questo ordinamento non è tanto giustificato. Per questo avevo deciso di fare più di un round, in modo tale che indipendentemente dall'ordinamento iniziale ogni feature set venga rivisto più volte.}
Such performance and the resulting order is reported, for each task, in Table~\ref{tab:featorder} in the Appendix.
%\label{sec:ordering}
%Once the initial candidate set has been selected, the feature blocks are processed sequentially (Section~\ref{sec:greedy}). The order in which the candidate feature blocks are inspected corresponds to their performance in isolation, and is reported %in~\ref{app:featureorder} (Table~\ref{tab:featorder}).
%in Table~\ref{tab:featorder}.
%\alemor{Fare qualche commento}

%--------------------------------------------------------------------

%\input{tab-selected-featuresV2}
\begin{table}[t]
    \small
    \centering
    \begin{tabular}{crccccc}
        \toprule
        &&& \multicolumn{3}{c}{\textbf{task}} \\
        \cmidrule{4-6}
        &&& \textit{activity} & \textit{toxicity} & \textit{diversity} \\
        \midrule
        \multicolumn{3}{r}{num. selected subgroups\;} & 22 (32\%) & 10 (14\%) & 25 (36\%) \\
        \midrule
        \multicolumn{3}{r}{MNMD \texttt{EMQ} (all features)\;} & 748.803 & 830.468 & 984.148 \\
        \multicolumn{3}{r}{MNMD \texttt{EMQO} (selected features)\;} & 619.986 & 626.650 & 787.219 \\
        \midrule
        \multicolumn{3}{r}{relative error reduction\;} & 17.20\% & 24.54\% & 20.01\% \\
        \bottomrule
    \end{tabular}
    \caption{Final selection of feature subgroups for each task. It reports the number of selected subgroups, the MNMD of the non-optimized EMQ models using all features, the MNMD of the EMQO models based on the selected subgroups, and the relative error reduction achieved by EMQO compared to EMQ.}
    \label{tab:selected-stats}
\end{table}

\paragraph{Selecting the Subgroups of Features} The greedy feature selection strategy described in Section~\ref{sec:greedy} is then applied to the ordering of Table~\ref{tab:featorder}. Figure~\ref{fig:feat-imp} shows the final results of the feature selection process where, for each task, the selected subgroups of features are represented by the green-coloured cells. In detail, the figure reports the percentage feature importance of each subgroup, where darker cells indicate higher importance, and blue cells denote that a subgroup was not chosen for the corresponding task. Additionally, Table~\ref{tab:selected-stats} provides summarized results per task, reporting the number of selected feature subgroups, the reference MNMD value of the non-optimized \texttt{EMQ} models that rely on all features, and the final MNMD corresponding to the \texttt{EMQO} models that make use of the selected feature subgroups. Finally, the relative error reduction of the \texttt{EMQO} models with respect to the non-optimized \texttt{EMQ} ones is reported in the bottommost row.
%\tizfag{Vi lascio una idea ma valutate voi se ne può valere la pena. Non so se è troppo lavoro ma mi stavo chiedendo se la tabella con le feature ottimali potesse anche essere integrata con la relativa importanza di quella feature nel task specifico. Ad esempio potremmo usare una heatmap per colourare ciascuna cella in base all'importanza di una feature in un task (ad esempio calcolando quanto il fatto di avere o non avere quella feature incida sull'errore complessivo del classificatore ottimale) e rendere più facilmemte comprensibile al lettore non solo quali sono le feature selezionate ma anche quali sono quelle più importanti (cosa che al momento mi sembra che manchi.).} \alemor{Volendo si puó fare, ma non so se viene più sintetico e chiaro com'è ora, cioè verde e rosso per dire scelta o non scelta. Poi secondo me l'informazione visuale che viene fuori è tanta che rimane un po' aneddotica; ci sará qualche caso simpatico da discutere e altri tanti scomodi da dover giustificare.}

\begin{figure*}[t]
 \centering
 \begin{subfigure}{0.4\textwidth}
 \includegraphics[width=1\linewidth]{ 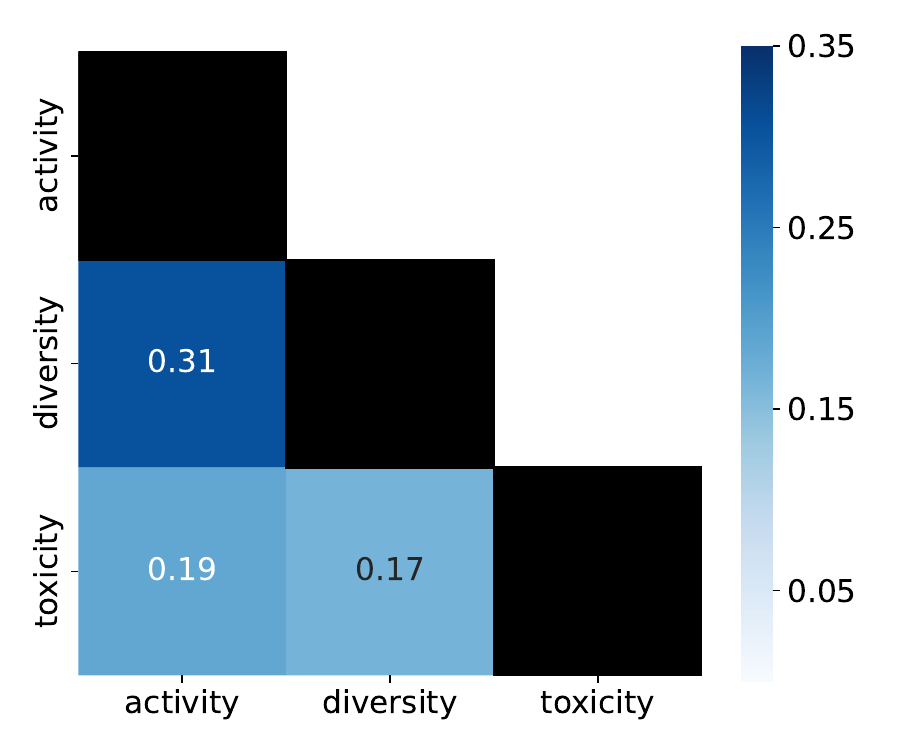}
 \centering
 \caption{Jaccard index.}
 \label{fig:overlap-jaccard}
 \end{subfigure}
 \hspace{0.05\textwidth}
 \begin{subfigure}{0.4\textwidth}
 \includegraphics[width=1\linewidth]{ 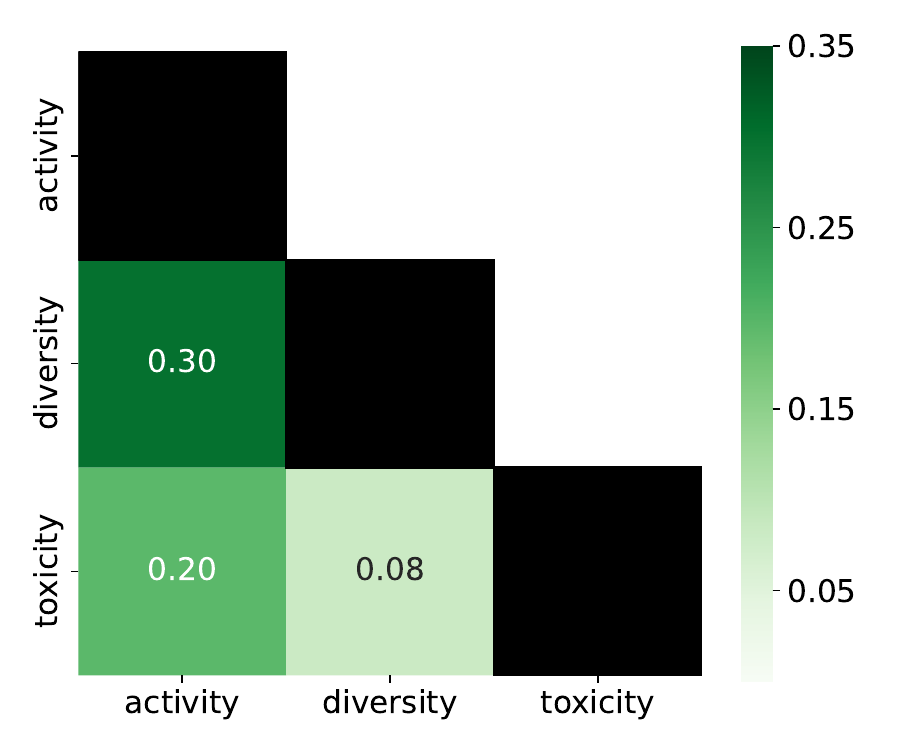}
 \centering
 \caption{Rank-Biased Overlap (RBO).}
 \label{fig:overlap-rbo}
 \end{subfigure}
 \caption{Pairwise similarity between the sets of features selected for the three behavioural prediction tasks: activity, toxicity, and diversity. Results are shown as heatmaps of the Jaccard index (left) and Rank-Biased Overlap (right). Larger values indicate greater similarity. Both measures reveal overall low similarity, highlighting that each task relies on largely distinct sets of predictive features.}
 \label{fig:overlap}
\end{figure*}

As already noticed in Figure~\ref{fig:curves}, aggregated results in Table~\ref{tab:selected-stats} reveal the \texttt{EMQO} models leveraging an optimized feature set markedly reduce the quantification error, with a reduction of 17.20\% in the activity task, 23.76\% in the toxicity task, and 18.15\% in the diversity one. %First, the number of selected feature subgroups varies considerably across tasks: only 22 (32\% of all available subgroups) for activity and 16 (23\%) for toxicity, compared to 31 (45\%) for diversity. This relatively compact selection for activity and toxicity suggests that a limited subset of features is sufficient to capture the behavioural changes in these dimensions, reflecting the stronger alignment between specific feature groups and the corresponding target variables.
Then, the number of selected feature subgroups varies considerably across tasks: only 10 (14\%) for toxicity compared to 22 (32\% of all available subgroups) for activity and 25 (36\%) for diversity. The relatively parsimonious feature selection for toxicity suggests that a limited subset of features is sufficient to capture the behavioural changes in this dimension. Concerning Figure~\ref{fig:feat-imp}, we notice that feature importance values are generally low and often close to zero, indicating a multitude of weakly informative features. Notable exceptions are the \texttt{TRENDS} subgroup that shows a large importance for the activity task and the \texttt{TOXICITY} subgroup that achieves a similar result for the toxicity task. These results align with the definitions of the tasks and are consistent with recent findings~\cite{tessa2025beyond}. In fact, the \texttt{TRENDS} subgroup alone achieves 19.8\% feature importance in the activity task, while the sum of the remaining features only accounts for 18.21\%. This dominance is reflected in a Gini coefficient $g=0.65$, indicating that this single subgroup largely outweighs the others. A similar pattern is observed for the toxicity task, where the \texttt{TOXICITY} subgroup achieves 11.4\% feature importance compared to 13.7\% of the remaining features summed together, with a Gini coefficient $g=0.53$. This shows that the \texttt{TOXICITY} subgroup is dominant in the namesake task, although less markedly so than the corresponding one in the activity task. In spite of the similar patterns however, the combined analysis of Figure~\ref{fig:feat-imp} and Table~\ref{tab:selected-stats} highlights some differences between the activity and toxicity tasks. In fact, despite using as much as 22 feature groups, the optimized activity model only improves on the non-optimized one by 17.20\%. This result suggests that beyond activity trends, the remaining features collectively provide little information for the activity task, contrary to the toxicity one whose improvement is 23.76\% with just 10 selected feature subgroups. Figure~\ref{fig:overlap} reports an analysis of the degree of similarity between the sets of features selected for the three tasks. We assess similarity using two complementary indicators. The Jaccard index $J$ captures the extent of overlap between feature sets, providing an intuitive measure of how many features are shared across tasks. Additionally, the Rank-Biased Overlap (RBO) considers the relative ordering of features, with higher weight given to those at the top of each ranking~\cite{webber2010similarity}. Both indicators provide values $\in [0,1]$, where 0 denotes no overlap and 1 perfect overlap. Results in Figure~\ref{fig:overlap} show weak similarity across tasks, with mean $J = 0.22$ and mean RBO $= 0.19$. Across both indicators, activity and diversity appear as the most similar tasks, while activity and toxicity as the most dissimilar. Overall, these results highlight the importance of designing specific rather than generic features for each task, and suggest that predicting possible changes in toxicity represents a simpler task than predicting changes in activity. Besides these results, diversity remains the most challenging task. This is reflected in Table~\ref{tab:selected-stats} by the largest set of selected features out of all tasks, and the largest quantification error for both the optimized and non-optimized models. Moreover, Figure~\ref{fig:feat-imp} shows that no subgroup of features stands out as significantly more important than the others, resulting in the lowest Gini coefficient $g=0.52$ out of the three tasks. These results highlight the intrinsic difficulty of the task as it is not straightforward to identify features that are strongly predictive of changes to participation diversity. Taken together, these findings suggest that predicting changes in participation diversity, and to a lower extent in activity, remains challenging with the available features, and is less amenable to performance gains through feature selection alone.

\section{Discussion and Conclusions}
\label{sec:Discussion}

We explored the ways in which a wide range of socio-behavioural, stylistic, and psychological features relate to user responses to a moderation intervention on Reddit. Our findings uncover several noteworthy patterns in how different types of user attributes correspond to distinct post-intervention behavioural changes.

\subsection{Consistently Informative and Uninformative Features}
Across all three behavioural prediction tasks---activity, toxicity, and participation diversity---a small set of staple feature groups consistently emerged as informative, i.e., \texttt{MENTIONS\_LINKS}, \texttt{TRENDS}, \texttt{PUNCTUATION}, and \texttt{TOXICITY}. Their persistent selection, regardless of the outcome being modelled, highlights their broad relevance in capturing user behaviour following a moderation intervention. For example, the inclusion of \texttt{TOXICITY} features in all tasks reflects the central role that toxic expression plays in shaping user responses on Reddit, especially in the context of the Great Ban, where affected communities were removed due to aggressive and harmful conduct.\footnote{\url{https://www.reddit.com/r/announcements/comments/hi3oht/update_to_our_content_policy/}} This suggests that toxic behaviour is not only predictive of future toxicity, but also intertwined with users’ overall engagement and community participation following a moderation intervention. Similarly, the consistent use of mention-related features points to the importance of social embeddedness---how users interact with others, reference communities, or cite external sources---as a general indicator of behavioural patterns. Activity and toxicity \texttt{TRENDS} also surfaced as widely useful, aligning with prior findings that noted the predictive power of such temporal dynamics in anticipating user reactions to moderation~\cite{tessa2025beyond}. Finally, the selection of \texttt{PUNCTUATION} features across all tasks supports recent work showing that punctuation is not merely structural, but often carries stylistic or pragmatic meaning---helping to signal tone, stance, or identity in online discourse~\cite{androutsopoulos2023punctuating}. Together, these features capture a blend of content, context, and expression that appears robustly predictive across multiple dimensions of user response to moderation.

While some feature groups proved consistently informative, a substantial portion (33\% of all feature groups) were never selected for any task. This outcome partly reflects the breadth of our approach, which deliberately incorporated a wide array of features drawn from recent literature on online behaviour and moderation~\cite{anjum2024hate, gilda2022predicting, raut2024enhancing, trujillo2022make}. Naturally, not all of these features are equally useful. Still, this result is informative in that it suggests that many features commonly used in related studies may offer limited value when it comes to predicting user reactions to moderation interventions. As an example, many of the features used herein were recently shown to boost online hate speech detection~\cite{raut2024enhancing}, a strictly related task. However, rather than dismissing these features outright, our results suggest that future work could benefit from prioritizing feature implementation by focusing on the groups and subgroups that show the most promise based on the task at hand. In any case, the overall sparsity of %Table~\ref{tab:selection}
Figure~\ref{fig:feat-imp}, where truly informative signals are relatively rare and for any given %behavioural dimension
task most available features contribute little to predictive performance, underscores the difficulty of the tasks. These findings highlight the complexity of predicting user responses to moderation~\cite{tessa2025beyond}, and underscore the need for continued scientific exploration into this practically important and socially impactful task.

\subsection{Task-Specific Feature Relevance}
When introducing our results in Section~\ref{sec:Results}, we noticed some relevant differences in predictive performance across the three behavioural tasks, with changes in activity and toxicity being estimated more accurately than changes in participation diversity. This gap in performance suggests that the diversity task might be inherently more challenging. Part of the difficulty could stem from the lack of a dedicated set of predictive features for the diversity task. In contrast, both the activity and toxicity tasks benefited from feature groups that were explicitly aligned with their respective behavioural dimensions, such as past activity levels and expressions of toxic language. This is relevant since our feature set reflects the current state of the art in modelling online behaviour~\cite{niverthi2022characterizing, raut2024enhancing, silva2021predicting}. It thus emerges that participation diversity represents a dimension of online user behaviour that has received so far limited attention compared to other more studied phenomena, despite diversity and plurality being widely recognized as important in a growing body of literature on online moderation and algorithmic fairness~\cite{fazelpour2022diversity, pedreschi2025human, trujillo2023one}. The lower performance on the diversity task highlights a research gap, and calls for further investigation into the behavioural signals that drive shifts in user participation.

Beyond differences in task difficulty, our analysis also revealed that the sets of features selected for each task %behavioural dimension 
are largely distinct. This suggests that predicting changes in activity, toxicity, and participation diversity, requires attending to different underlying drivers of user behaviour. For example, \texttt{RELATIONAL} features---capturing interactions with other users and communities---were almost entirely absent in the selected sets for the activity and toxicity tasks. Yet all relational subgroups were included for the diversity task. Indeed, participation diversity is inherently related to the structure and breadth of a user’s engagement across subreddits. \texttt{WRITING\_STYLE} features also emerged as more relevant to diversity: 60\% of writing style -related subgroups were selected for this task, compared to only 25\% and 20\% for activity and toxicity, respectively. This observation resonates with recent findings that different communities often exhibit distinctive linguistic norms~\cite{shen2022tale, trujillo2023one}, suggesting that variation in writing style may reflect the diversity of contexts in which a user participates. Such features, therefore, offer indirect signals about a user’s exposure to and integration within varied online spaces. Overall, these findings emphasize the importance of tailoring feature design to the specific behavioural dimension under study, and point toward more nuanced, context-aware modelling approaches for capturing the multifaceted ways users respond to moderation~\cite{cima2025contextualized, cresci2022personalized}. Other than this, our results also have practical implications regarding the generalizability of predictive models in this domain. Our findings suggest that distinct aspects of user behaviour, such as activity levels, toxicity, or participation diversity, are influenced by different sets of features, and therefore require dedicated predictors. This might limit the feasibility of using a single, unified model to estimate the overall impact of a moderation intervention. Instead, accurately capturing the full spectrum of possible user responses might require training and maintaining separate predictive systems for each behavioural outcome of interest, which may complicate large-scale or real-time applications.

\subsection{Toward Objective-Specific Moderation Strategies}
The results presented in %Table~\ref{tab:selection}
Figure~\ref{fig:feat-imp} highlighted the nuanced role that different user features play in predicting behavioural changes following moderation interventions. While some feature subgroups were consistently selected across tasks and others were never selected, these two extremes account for only about 40\% of all considered subgroups. The majority of features fall into an intermediate space, i.e., they are predictive for some outcomes but not others. This variability suggests that user reactions to moderation are multifaceted and %depend on \fabseb{Secondo me qui non è un ``depend on'' ma un ``manifest''.}
manifest distinct behavioural signals depending on the dimension being targeted. This finding carries implications for how moderation strategies are designed and deployed. For instance, interventions aimed at encouraging broader participation across communities---such as efforts to reduce echo chambers or increase exposure to diverse viewpoints---should be grounded in the behavioural signals most relevant to participation diversity. Conversely, strategies targeting reductions in toxic behaviour should be informed by a different set of user characteristics. In other words, our results point to the need for \textit{objective-specific moderation}, where each intervention is tailored not only to the user it targets, but also to the specific behavioural outcome it seeks to influence.

This perspective adds a new layer to the growing call for more user-centred and personalized moderation approaches. While prior work has emphasized that moderation should consider the traits and histories of the users involved~\cite{drazkiewicz2022study, cresci2022personalized}, our findings complement and extend this view by arguing that the design of interventions should also be guided by the nature of the intended effect. A one-size-fits-all approach, such as blanket bans applied indiscriminately across cases, may be ill-suited to achieve diverse moderation goals. In contrast, a more principled, data-driven strategy that aligns intervention types with the behavioural changes they aim to foster offers a promising path forward. In a content moderation landscape that continues to rely heavily on a narrow set of interventions~\cite{trujillo2025dsa, shahi2025year}, our findings underscore the importance of expanding both the methodological and conceptual toolkits used to design them. Tailoring interventions not just to who the user is, but to what the intervention is meant to accomplish, represents a favourable step toward effective, fair, and goal-sensitive moderation.

\subsection{Limitations and Future Work}
Our analysis focuses on Reddit’s 2020 Great Ban, a large-scale deplatforming intervention~\cite{cima2024great}. Even if such intervention affected a large number of users and communities, the reliance on a single type of moderation (i.e., a community ban) at a single point in time, nonetheless represents a limitation of our work. Similarly, our results are derived solely from Reddit data, a platform with a large user base, but with specific structural and cultural characteristics. Together, these two factors may limit the generalizability of our findings to other moderation contexts or platforms, though such constraints are standard in empirical computational research on content moderation. Our results are also based on the set of quantification algorithms and user features we experimented with. While we selected well-established methods and drew on an extensive feature set, alternative algorithms or unexplored features could yield additional insights. Moreover, our models identify correlational patterns between user attributes and post-intervention behaviour, but do not support causal claims. Determining whether specific features cause users to react in certain ways would require dedicated experimental or quasi-experimental designs that could be the goal of future research in this domain. Finally, our approach does not incorporate temporal or longitudinal modelling, which presents a further promising direction for future work.

Adding to the previous research directions, our results pave the way for future predictive systems capable of accurately modelling user reactions to moderation, whether via classification, quantification, or regression approaches. A parallel research avenue involves extending this analysis to other forms of moderation and additional platforms, though progress in this direction may be constrained by the limited accessibility of platform data~\cite{assenmacher2021benchmarking}. Towards the medium or long term, our work also sets the stage for the development of moderation strategies that are both context-aware and tailored to specific intervention goals, marking a step toward more nuanced and effective content moderation frameworks.

%--------------------------------------------------------------------

%\section{Conclusions}
%\label{sec:Conclusions}

%--------------------------------------------------------------------

\section*{Acknowledgments}
\label{sec:Acknowledgments}
\ifdraftelse{
% This work has been funded by the QuaDaSh project \enquote{Finanziato dall’Unione europea---Next Generation EU, Missione 4 Componente 2 CUP B53D23026250001}.
% }{
 AM's and FS's work has been partially supported by project ``Future Artificial Intelligence Research'' (FAIR -- CUP B53D22000980006), project ``Quantification under Dataset Shift'' (QuaDaSh -- CUP B53D23026250001), and project ``Strengthening the Italian RI for Social Mining and Big Data Analytics'' (SoBigData.it -- CUP B53C22001760006), all funded by the European Union under the NextGenerationEU funding scheme. SC's, BT's, and TF's work has been partly supported by the European Union -- Next Generation EU, Mission 4 Component 1, for project PIANO (CUP B53D23013290006) and for the ERC project DEDUCE under grant \#101113826.
}

%--------------------------------------------------------------------

\bibliography{Benedetta, Fabrizio, Stefano, Tiziano}
\bibliographystyle{plain}

%--------------------------------------------------------------------

\appendix
\glsaddall
\pagebreak
%\printglossary
\section{Features}
Table~\ref{tab:features_split} reports the cardinality of each subgroup of features, while Table~\ref{tab:glossary} provides their description.

\begin{table}[ht!]
%\centering
\begin{minipage}{0.48\textwidth}
\footnotesize
\begin{tabular}{lll}
\textbf{group} & \textbf{subgroup} & \textbf{num} \\
\toprule
\texttt{EMBEDDINGS} & \texttt{EMBEDDINGS} & \texttt{384} \\
\midrule
\texttt{LIWC} & \texttt{LINGUISTIC} & \texttt{21} \\
             & \texttt{SOCIAL} & \texttt{12} \\
             & \texttt{COGNITION} & \texttt{12} \\
             & \texttt{AFFECT} & \texttt{11} \\
             & \texttt{MOTIVATION} & \texttt{9} \\
             & \texttt{SUMMARY} & \texttt{8} \\
             & \texttt{LIFESTYLE} & \texttt{6} \\
             & \texttt{BIO} & \texttt{6} \\
             & \texttt{DRIVES} & \texttt{4} \\
             & \texttt{CULTURE} & \texttt{4} \\
             & \texttt{SPATIAL} & \texttt{4} \\
             & \texttt{TEMPORAL} & \texttt{4} \\
             & \texttt{CONVERSATION} & \texttt{4} \\
\midrule
\texttt{WRITING\_STYLE} & \texttt{NER} & \texttt{5} \\
             & \texttt{FLESCH-KINKAID} & \texttt{4} \\
             & \texttt{SMOG} & \texttt{4} \\
             & \texttt{SPELL\_ERRORS} & \texttt{4} \\
             & \texttt{NOUNS} & \texttt{4} \\
             & \texttt{ARTICLES} & \texttt{4} \\
             & \texttt{ADPOSITIONS} & \texttt{4} \\
             & \texttt{IRONY} & \texttt{4} \\
             & \texttt{NOVELTY} & \texttt{4} \\
             & \texttt{STOPWORDS} & \texttt{4} \\
             \midrule
             \texttt{WRITING\_STYLE} & \texttt{SYM} & \texttt{4} \\
             & \texttt{CONG} & \texttt{4} \\
             & \texttt{SCONJ} & \texttt{4} \\
             & \texttt{PROPN} & \texttt{4} \\
             & \texttt{PRON} & \texttt{4} \\
             & \texttt{INTJ} & \texttt{4} \\
             & \texttt{DET} & \texttt{4} \\
             & \texttt{ADV} & \texttt{4} \\
             & \texttt{VERB} & \texttt{4} \\
             & \texttt{AUX} & \texttt{4} \\

\bottomrule
\end{tabular}
\end{minipage}%
\hfill
\begin{minipage}{0.48\textwidth}
\footnotesize
\centering
\begin{tabular}{lll}
\textbf{group} & \textbf{subgroup} & \textbf{num} \\
\toprule
\texttt{ACTIVITY} & \texttt{SUBMISSIONS} & \texttt{14} \\
             & \texttt{COMMENTS} & \texttt{9} \\
             & \texttt{TIME} & \texttt{6} \\
             & \texttt{MENTIONS\_LINKS} & \texttt{3} \\
             & \texttt{TRENDS} & \texttt{2} \\
         \midrule
\texttt{RELATIONAL} & \texttt{ENGAGEMENT} & \texttt{10} \\
             & \texttt{THREADS} & \texttt{10} \\
             & \texttt{AUTHORS} & \texttt{10} \\
             & \texttt{SUBR} & \texttt{6} \\
             \midrule
\texttt{SENTIMENT} & \texttt{NEG} & \texttt{4} \\
             & \texttt{NEU} & \texttt{4} \\
             & \texttt{POS} & \texttt{4} \\
             & \texttt{CPD} & \texttt{4} \\
             & \texttt{NEG-EMOJI} & \texttt{4} \\
             & \texttt{NEU-EMOJI} & \texttt{4} \\
             & \texttt{POS-EMOJI} & \texttt{4} \\
             & \texttt{SENT-EMOJI} & \texttt{4} \\
             \midrule
\texttt{SOC\_PSY} & \texttt{OCEAN} & \texttt{5} \\
             & \texttt{AUTHORITY} & \texttt{4} \\
             & \texttt{FAIRNESS} & \texttt{4} \\
             & \texttt{GENERAL} & \texttt{4} \\
             & \texttt{HARM} & \texttt{4} \\
             & \texttt{INGROUP} & \texttt{4} \\
             & \texttt{PURITY} & \texttt{4} \\
             & \texttt{DEMOGRAPHIC} & \texttt{2} \\
             \midrule
\texttt{TOXICITY} & \texttt{TOXICITY} & \texttt{4} \\
             & \texttt{SEVERE-TOXICITY} & \texttt{4} \\
             & \texttt{OBSCENE} & \texttt{4} \\
             & \texttt{THREAT} & \texttt{4} \\
             & \texttt{INSULT} & \texttt{4} \\
             & \texttt{ID-ATTACK} & \texttt{4} \\
             \midrule
\texttt{EMOTIONS} & \texttt{EIL} & \texttt{8} \\
             & \texttt{EMOSCORES} & \texttt{8} \\
             & \texttt{VAD} & \texttt{3} \\
\bottomrule
\end{tabular}
\end{minipage}
\caption{Feature groups, subgroups, and their cardinality.}
\label{tab:features_split}
\end{table}

 \footnotesize
\begin{longtable}[t!]{clp{0.6\textwidth}}
\toprule
\textbf{Group} & \textbf{Subgroup} & \textbf{Description} \\ 
\midrule
\endfirsthead
\multicolumn{3}{c}{\textit{Table~\ref{tab:glossary} (continued from previous page)}}\\
\toprule
\textbf{Group} & \textbf{Subgroup} & \textbf{Description} \\ 
\midrule
\endhead
        % Footer for first page
        \midrule
        \multicolumn{3}{c}{\textit{Table~\ref{tab:glossary} continues on next page}} \\
        \bottomrule
        \endfoot
        % Final footer
        \endlastfoot
\multirow{1}{*}{\rotatebox{90}{\texttt{EMB}}} & \texttt{EMBEDDINGS} & Numerical vector representations of text generated using a sentence-transformer \\ 
\midrule
\multirow{23}{*}{\rotatebox[origin=c]{90}{\texttt{LIWC}}}
  & \texttt{LINGUISTIC} 
  & Linguistic features such as pronouns, articles, function words, and verb tenses \\
 &  \texttt{SOCIAL} & Words that reflect social processes and relationships, including terms related to family, friends, and people \\ 
 &  \texttt{COGNITION} & Vocabulary linked to cognitive functions such as insight, causation, certainty, and tentativeness \\ 
 & \texttt{AFFECT} & Terms expressing affective states like positive and negative emotions, anxiety, anger, and sadness \\ 
 &  \texttt{MOTIVATION} & Words related to drives and needs, such as achievement, power, reward, and affiliation \\ 
 &  \texttt{SUMMARY} & Aggregate measures computed by LIWC, such as analytical thinking, clout, authenticity, and emotional tone \\ 
 &  \texttt{LIFESTYLE} & Terms associated with everyday life domains, such as work, home, money, and leisure activities \\ 
 &  \texttt{BIO} & Words referring to biological processes, including health, body, and sexual content \\ 
 &  \texttt{DRIVES} & Lexical indicators of psychological drives such as affiliation, achievement, and risk \\ 
 &  \texttt{CULTURE} & References to cultural terms including religion, morality, and traditions \\ 
 &  \texttt{SPATIAL} & Words indicating spatial relations and references to locations or movements \\ 
 &  \texttt{TEMPORAL} & Expressions related to time, including past, present, and future references \\ 
 &  \texttt{CONVERSATION} & Lexical items typically used in conversations, such as assent, fillers, and nonfluencies \\ 
\midrule
\multirow{12}{*}{\rotatebox[origin=c]{90}{\texttt{SENTIMENT}}} &  \texttt{NEGATIVE} & Negative sentiment expressed in text, capturing feelings such as sadness, anger, or dissatisfaction \\ 
 &  \texttt{NEUTRAL} & Neutral or objective tone without strong emotional charge \\ 
 &  \texttt{POSITIVE} & Positive sentiment reflected in language conveying happiness, approval, or optimism \\ 
 &  \texttt{COMPOUND} & Overall sentiment score combining positive and negative expressions into a single value \\ 
 &  \texttt{NEGATIVE EMOJI} & Use of emojis conveying negative emotions such as sadness, anger, or frustration \\ 
 &  \texttt{NEUTRAL EMOJI} & Use of emojis expressing neutral or ambiguous emotions \\ 
 &  \texttt{POSITIVE EMOJI} & Use of emojis conveying positive emotions like happiness, love, or encouragement \\ 
 &  \texttt{SENTIMENT EMOJI} & Overall sentiment expressed through emojis combining positive, neutral, and negative cues \\ 
\midrule
 \multirow{2}{*}{\rotatebox{90}{\texttt{TOXICITY}}} & \texttt{TOXICITY} & Presence of toxic language, including insults, offensive words, or harmful content \\ 
 &  \texttt{SEVERE TOXICITY} & Strong forms of toxic content characterized by extreme insults or threats \\ 
 \multirow{6}{*}{\rotatebox{90}{\texttt{TOXICITY}}} &  \texttt{OBSCENE} & Use of obscene or vulgar language and expressions \\ 
 &  \texttt{THREAT} & Statements containing threats of harm or violence \\ 
 &  \texttt{INSULT} & Language intended to offend or demean another individual or group \\ 
 &  \texttt{IDENTITY ATTACK} & Offensive comments targeting a person's identity attributes such as race, gender, or religion \\ 
\midrule
\multirow{14}{*}{\rotatebox{90}{ \texttt{SOC\_PSY}}} & \texttt{OCEAN} & Personality traits based on the Big Five model: Openness, Conscientiousness, Extraversion, Agreeableness, and Neuroticism \\ 
 & \texttt{AUTHORITY} & Words related to respect for authority, hierarchy, and social order \\ 
 & \texttt{FAIRNESS} & Expressions concerning justice, equity, and fairness in social contexts \\ 
 & \texttt{GENERAL} & General social psychological terms that do not fit into more specific categories \\ 
&  \texttt{HARM} & Terms related to harm, suffering, or protection from harm in social situations \\ 
 & \texttt{INGROUP} & Words denoting belonging to or identification with social groups \\ 
 & \texttt{PURITY} & Language reflecting concepts of purity, sanctity, and moral cleanliness \\ 
 &  \texttt{DEMOGRAPHIC} & Terms related to demographic characteristics such as age, gender, or ethnicity \\ \midrule 
\multirow{8}{*}{\rotatebox{90}{\texttt{ACTIVITY}}} &  \texttt{SUBMISSIONS} & Information about users' submissions made on Reddit \\ 
 &  \texttt{COMMENTS} & Information about users' comments made on Reddit \\ 
 &  \texttt{TIME} & Temporal aspects of user activity, including frequency and timing of posts  \\ 
 &  \texttt{MENTIONS\_LINKS} & Number of comments where users mention other users, communities or include hyperlinks in their content \\ 
 &  \texttt{TRENDS} & Slope representing the increases or decreases in activity and toxicity \\ 
\midrule
\multirow{6}{*}{\rotatebox{90}{\texttt{RELATIONAL}}} &  \texttt{ENGAGEMENT} & Measures of user interaction computed as the difference between upvotes and downvotes to a comment \\ 
 &  \texttt{THREADS} & Conversation chains or discussion threads where users participate \\ 
 &  \texttt{AUTHORS} & Information about how much users interact with other users \\ 
 &  \texttt{SUBREDDITS} & Data related to the specific subreddit communities where users post or comment \\ 
 \midrule
\multirow{6}{*}{\rotatebox{90}{\texttt{EMOTIONS}}} & \texttt{EIL} & Categories of basic emotions (anger, anticipation, disgust, fear, joy, sadness, surprise, trust) identified using the EIL lexicon \\ 

 & \texttt{EMOSCORES} & Numerical scores quantifying the intensity of emotions (anger, trust, surprise, disgust, joy, sadness, fear, anticipation) \\ 

 & \texttt{VAD} & Dimensions of affective meaning: Valence (positive–negative), Arousal (calm–excited), Dominance (submissive–dominant) \\ 
\midrule
\multirow{5}{*}{\rotatebox{90}{\texttt{WR\_STYLE}}} & \texttt{FLESCH-KINKAID} & Readability index based on the flesch-kinkaid score \\ 

 & \texttt{SMOG} & Readability index estimating years of education required to understand text \\ 

 & \texttt{SPELL\_ERRORS} & Frequency of spelling mistakes in text \\ 

 & \texttt{VERBS} & Distribution of verbs used in comments \\ 

\multirow{15}{*}{\rotatebox{90}{\texttt{WRITING\_STYLE}}} & \texttt{NOUNS} & Distribution of nouns used in comments \\ 

 & \texttt{ARTICLES} & Distribution of articles (e.g., “the”, “a”) used in comments \\ 

 & \texttt{ADVERBS} & Distribution of adverbs used in comments \\ 

 & \texttt{ADPOSITIONS} & Distribution of adpositions (e.g., prepositions, postpositions) used in comments \\ 

 & \texttt{PRONOUNS} & Distribution of pronouns used in comments \\ 

 & \texttt{IRONY} & Estimated presence of ironic expressions in text \\ 

 & \texttt{NER} & Named entity recognition categories (e.g., persons, organizations, geopolitical entities, laws, groups) \\ 

& \texttt{NOVELTY} & Measure of textual originality compared to reference corpus \\ 

 & \texttt{STOPWORDS} & Distribution of stopwords (common non-content words such as “and”, “the”, “is”) \\ 

& \texttt{SYM} & Distribution of symbols present in text (e.g., \%, \$, \@) \\ 

 & \texttt{CONG} & Distribution of coordinating conjunctions (e.g., “and”, “but”) \\ 

 & \texttt{SCONJ} & Distribution of subordinating conjunctions (e.g., “because”, “although”) \\ 

 & \texttt{PROPN} & Distribution of proper nouns in text \\ 

& \texttt{PRON} & Distribution of pronouns in text \\ 

& \texttt{INTJ} & Distribution of interjections (e.g., “wow”, “oh”) \\ 

& \texttt{DET} & Distribution of determiners (e.g., “this”, “those”) \\ 

& \texttt{ADV} & Distribution of adverbs (fine-grained category from POS tagging) \\ 

 & \texttt{VERB} & Distribution of verbs (fine-grained category from POS tagging) \\ 

 & \texttt{AUX} & Distribution of auxiliary verbs (e.g., “is”, “have”, “will”) \\ 
\bottomrule
\caption{Brief description of each subgroup of features.}
\label{tab:glossary}
\end{longtable}

%\label{app:featuretable}

\section{Quantification Performance of the Subgroups of Features}
Table~\ref{tab:featorder} reports the order of inspection of the feature subgroups for each task. This order is used during the greedy feature selection process described in Algorithm~\ref{alg:greedy}.
\begin{center}
    \footnotesize
    \setlength{\tabcolsep}{3pt}
    %\adjustbox{max width=\textwidth}{
    \begin{longtable}[t]{crrrrrr}
        % header
        \toprule
        & \multicolumn{2}{c}{\textbf{activity}} & \multicolumn{2}{c}{\textbf{toxicity}} & \multicolumn{2}{c}{\textbf{diversity}} \\
        \cmidrule(rl){2-3} \cmidrule(rl){4-5} \cmidrule(rl){6-7}
        \textbf{rank} & \multicolumn{1}{c}{feature name} & \multicolumn{1}{c}{MNMD} & \multicolumn{1}{c}{feature name} & \multicolumn{1}{c}{MNMD} & \multicolumn{1}{c}{feature name} & \multicolumn{1}{c}{MNMD} \\
        \midrule
        \endfirsthead
        % Continuation header for second page
        \multicolumn{7}{c}{\textit{Table~\ref{tab:featorder} (continued from previous page)}} \\
        \toprule
        & \multicolumn{2}{c}{\textbf{activity}} & \multicolumn{2}{c}{\textbf{toxicity}} & \multicolumn{2}{c}{\textbf{diversity}} \\
        \cmidrule(rl){2-3} \cmidrule(rl){4-5} \cmidrule(rl){6-7}
        \textbf{rank} & \multicolumn{1}{c}{feature name} & \multicolumn{1}{c}{MNMD} & \multicolumn{1}{c}{feature name} & \multicolumn{1}{c}{MNMD} & \multicolumn{1}{c}{feature name} & \multicolumn{1}{c}{MNMD} \\
        \midrule
        \endhead
        % Footer for first page
        \midrule
        \multicolumn{7}{c}{\textit{Table~\ref{tab:featorder} continues on next page}} \\
        \bottomrule
        \endfoot
        % Final footer
        \endlastfoot
1 & \cellcolor{lightblue}\texttt{SOC:AUTHORITY} & \cellcolor{lightblue}1657.735 & \cellcolor{lightgreen}\texttt{REL:ENGAGEMENT} & \cellcolor{lightgreen}1368.431 & \cellcolor{lightblue}\texttt{SOC:AUTHORITY} & \cellcolor{lightblue}1311.063 \\
2 & \cellcolor{lightyellow}\texttt{LIW:CULTURE} & \cellcolor{lightyellow}1644.752 & \cellcolor{lightgreen}\texttt{REL:AUTHORS} & \cellcolor{lightgreen}1366.133 & \cellcolor{lightgreen}\texttt{REL:AUTHORS} & \cellcolor{lightgreen}1271.090 \\
3 & \cellcolor{lightteal}\texttt{WRI:CONG} & \cellcolor{lightteal}1599.881 & \cellcolor{lightyellow}\texttt{LIW:COGNITION} & \cellcolor{lightyellow}1328.143 & \cellcolor{lightblue}\texttt{SOC:FAIRNESS} & \cellcolor{lightblue}1265.083 \\
4 & \cellcolor{lightteal}\texttt{WRI:SYM} & \cellcolor{lightteal}1585.881 & \cellcolor{lightgreen}\texttt{REL:THREADS} & \cellcolor{lightgreen}1316.311 & \cellcolor{lightyellow}\texttt{LIW:SUMMARY} & \cellcolor{lightyellow}1262.627 \\
5 & \cellcolor{lightyellow}\texttt{LIW:DRIVES} & \cellcolor{lightyellow}1583.815 & \cellcolor{lightyellow}\texttt{LIW:BIO} & \cellcolor{lightyellow}1312.633 & \cellcolor{lightyellow}\texttt{LIW:COGNITION} & \cellcolor{lightyellow}1234.172 \\
6 & \cellcolor{lightblue}\texttt{SOC:HARM} & \cellcolor{lightblue}1564.193 & \cellcolor{lightyellow}\texttt{LIW:PUNCTUATION} & \cellcolor{lightyellow}1311.661 & \cellcolor{lightyellow}\texttt{LIW:LINGUISTIC} & \cellcolor{lightyellow}1231.092 \\
7 & \cellcolor{lightorange}\texttt{ACT:MENTIONS\_LINKS} & \cellcolor{lightorange}1557.762 & \cellcolor{lightred}\texttt{EMO:EIL} & \cellcolor{lightred}1279.451 & \cellcolor{lightteal}\texttt{WRI:SMOG} & \cellcolor{lightteal}1227.331 \\
8 & \cellcolor{lightyellow}\texttt{LIW:MOTIVATION} & \cellcolor{lightyellow}1540.192 & \cellcolor{lightblue}\texttt{SOC:DEMOGRAPHIC} & \cellcolor{lightblue}1270.266 & \cellcolor{lightteal}\texttt{WRI:PRON} & \cellcolor{lightteal}1224.617 \\
9 & \cellcolor{lightred}\texttt{EMO:VAD} & \cellcolor{lightred}1520.968 & \cellcolor{lightred}\texttt{EMO:EMOSCORES} & \cellcolor{lightred}1260.259 & \cellcolor{lightgreen}\texttt{REL:THREADS} & \cellcolor{lightgreen}1215.066 \\
10 & \cellcolor{lightyellow}\texttt{LIW:PUNCTUATION} & \cellcolor{lightyellow}1515.679 & \cellcolor{lightorange}\texttt{ACT:COMMENTS} & \cellcolor{lightorange}1254.685 & \cellcolor{lightred}\texttt{EMO:EIL} & \cellcolor{lightred}1205.383 \\
11 & \cellcolor{lightyellow}\texttt{LIW:TEMPORAL} & \cellcolor{lightyellow}1514.538 & \cellcolor{lightyellow}\texttt{LIW:LINGUISTIC} & \cellcolor{lightyellow}1241.606 & \cellcolor{lightorange}\texttt{ACT:SUBMISSIONS} & \cellcolor{lightorange}1205.228 \\
12 & \cellcolor{lightred}\texttt{EMO:EIL} & \cellcolor{lightred}1508.496 & \cellcolor{lightpurple}\texttt{SEN:POS-EMOJI} & \cellcolor{lightpurple}1220.949 & \cellcolor{lightteal}\texttt{WRI:FLESCH-KINKAID} & \cellcolor{lightteal}1198.309 \\
13 & \cellcolor{lightblue}\texttt{SOC:OCEAN} & \cellcolor{lightblue}1498.040 & \cellcolor{lightyellow}\texttt{LIW:SOCIAL} & \cellcolor{lightyellow}1218.462 & \cellcolor{lightteal}\texttt{WRI:NER} & \cellcolor{lightteal}1195.571 \\
14 & \cellcolor{lightpurple}\texttt{SEN:NEU-EMOJI} & \cellcolor{lightpurple}1475.566 & \cellcolor{lightyellow}\texttt{LIW:CONVERSATION} & \cellcolor{lightyellow}1212.394 & \cellcolor{lightblue}\texttt{SOC:PURITY} & \cellcolor{lightblue}1190.555 \\
15 & \cellcolor{lightblue}\texttt{SOC:INGROUP} & \cellcolor{lightblue}1469.145 & \cellcolor{lightorange}\texttt{ACT:SUBMISSIONS} & \cellcolor{lightorange}1212.151 & \cellcolor{lightyellow}\texttt{LIW:DRIVES} & \cellcolor{lightyellow}1185.555 \\
16 & \cellcolor{lightpurple}\texttt{SEN:SENT-EMOJI} & \cellcolor{lightpurple}1469.101 & \cellcolor{lightteal}\texttt{WRI:NOVELTY} & \cellcolor{lightteal}1211.304 & \cellcolor{lightorange}\texttt{ACT:TRENDS} & \cellcolor{lightorange}1181.687 \\
17 & \cellcolor{lightpurple}\texttt{SEN:POS-EMOJI} & \cellcolor{lightpurple}1464.278 & \cellcolor{lightteal}\texttt{WRI:NER} & \cellcolor{lightteal}1209.736 & \cellcolor{lightgreen}\texttt{REL:ENGAGEMENT} & \cellcolor{lightgreen}1179.924 \\
18 & \cellcolor{lightteal}\texttt{WRI:AUX} & \cellcolor{lightteal}1445.009 & \cellcolor{lightorange}\texttt{ACT:TIME} & \cellcolor{lightorange}1208.817 & \cellcolor{lightpurple}\texttt{SEN:NEG} & \cellcolor{lightpurple}1174.586 \\
19 & \cellcolor{lightpurple}\texttt{SEN:NEG-EMOJI} & \cellcolor{lightpurple}1443.660 & \cellcolor{lightblue}\texttt{SOC:GENERAL} & \cellcolor{lightblue}1205.825 & \cellcolor{lightteal}\texttt{WRI:SCONJ} & \cellcolor{lightteal}1168.685 \\
20 & \cellcolor{lightteal}\texttt{WRI:SCONJ} & \cellcolor{lightteal}1436.706 & \cellcolor{lightblue}\texttt{SOC:OCEAN} & \cellcolor{lightblue}1204.764 & \cellcolor{lightorange}\texttt{ACT:COMMENTS} & \cellcolor{lightorange}1168.595 \\
21 & \cellcolor{lightyellow}\texttt{LIW:SPATIAL} & \cellcolor{lightyellow}1430.364 & \cellcolor{lightteal}\texttt{WRI:DET} & \cellcolor{lightteal}1201.019 & \cellcolor{lightblue}\texttt{SOC:GENERAL} & \cellcolor{lightblue}1167.661 \\
22 & \cellcolor{lightorange}\texttt{ACT:TIME} & \cellcolor{lightorange}1425.299 & \cellcolor{lightyellow}\texttt{LIW:MOTIVATION} & \cellcolor{lightyellow}1199.623 & \cellcolor{lightteal}\texttt{WRI:SYM} & \cellcolor{lightteal}1162.521 \\
23 & \cellcolor{lightpink}\texttt{TOX:SEVERE-TOXICITY} & \cellcolor{lightpink}1420.325 & \cellcolor{lightyellow}\texttt{LIW:LIFESTYLE} & \cellcolor{lightyellow}1195.578 & \cellcolor{lightblue}\texttt{SOC:OCEAN} & \cellcolor{lightblue}1158.915 \\
24 & \cellcolor{lightyellow}\texttt{LIW:LIFESTYLE} & \cellcolor{lightyellow}1413.914 & \cellcolor{lightyellow}\texttt{LIW:AFFECT} & \cellcolor{lightyellow}1194.761 & \cellcolor{lightpink}\texttt{TOX:ID-ATTACK} & \cellcolor{lightpink}1158.599 \\
25 & \cellcolor{lightblue}\texttt{SOC:DEMOGRAPHIC} & \cellcolor{lightblue}1408.487 & \cellcolor{lightyellow}\texttt{LIW:SPATIAL} & \cellcolor{lightyellow}1190.505 & \cellcolor{lightred}\texttt{EMO:VAD} & \cellcolor{lightred}1158.457 \\
26 & \cellcolor{lightteal}\texttt{WRI:FLESCH-KINKAID} & \cellcolor{lightteal}1381.244 & \cellcolor{lightred}\texttt{EMO:VAD} & \cellcolor{lightred}1190.196 & \cellcolor{lightred}\texttt{EMO:EMOSCORES} & \cellcolor{lightred}1158.257 \\
27 & \cellcolor{lightteal}\texttt{WRI:NER} & \cellcolor{lightteal}1377.075 & \cellcolor{lightteal}\texttt{WRI:STOPWORDS} & \cellcolor{lightteal}1187.248 & \cellcolor{lightyellow}\texttt{LIW:SOCIAL} & \cellcolor{lightyellow}1156.770 \\
28 & \cellcolor{lightred}\texttt{EMO:EMOSCORES} & \cellcolor{lightred}1375.618 & \cellcolor{lightyellow}\texttt{LIW:CULTURE} & \cellcolor{lightyellow}1187.248 & \cellcolor{lightblue}\texttt{SOC:INGROUP} & \cellcolor{lightblue}1156.378 \\
29 & \cellcolor{lightpink}\texttt{TOX:OBSCENE} & \cellcolor{lightpink}1375.335 & \cellcolor{lightpurple}\texttt{SEN:NEU-EMOJI} & \cellcolor{lightpurple}1185.784 & \cellcolor{lightyellow}\texttt{LIW:BIO} & \cellcolor{lightyellow}1155.709 \\
30 & \cellcolor{lightpurple}\texttt{SEN:POS} & \cellcolor{lightpurple}1368.405 & \cellcolor{lightgreen}\texttt{REL:SUBR} & \cellcolor{lightgreen}1184.372 & \cellcolor{lightpink}\texttt{TOX:THREAT} & \cellcolor{lightpink}1155.021 \\
31 & \cellcolor{lightpurple}\texttt{SEN:CPD} & \cellcolor{lightpurple}1365.942 & \cellcolor{lightteal}\texttt{WRI:ARTICLES} & \cellcolor{lightteal}1182.534 & \cellcolor{lightteal}\texttt{WRI:AUX} & \cellcolor{lightteal}1152.659 \\
32 & \cellcolor{lightyellow}\texttt{LIW:COGNITION} & \cellcolor{lightyellow}1365.580 & \cellcolor{lightteal}\texttt{WRI:SYM} & \cellcolor{lightteal}1180.736 & \cellcolor{lightorange}\texttt{ACT:TIME} & \cellcolor{lightorange}1151.716 \\
33 & \cellcolor{lightteal}\texttt{WRI:SMOG} & \cellcolor{lightteal}1361.447 & \cellcolor{lightpurple}\texttt{SEN:NEG-EMOJI} & \cellcolor{lightpurple}1178.618 & \cellcolor{lightteal}\texttt{WRI:ARTICLES} & \cellcolor{lightteal}1150.436 \\
34 & \cellcolor{lightblue}\texttt{SOC:FAIRNESS} & \cellcolor{lightblue}1360.234 & \cellcolor{lightteal}\texttt{WRI:NOUNS} & \cellcolor{lightteal}1177.560 & \cellcolor{lightblue}\texttt{SOC:HARM} & \cellcolor{lightblue}1149.303 \\
35 & \cellcolor{lightyellow}\texttt{LIW:LINGUISTIC} & \cellcolor{lightyellow}1357.386 & \cellcolor{lightteal}\texttt{WRI:ADV} & \cellcolor{lightteal}1175.680 & \cellcolor{lightyellow}\texttt{LIW:PUNCTUATION} & \cellcolor{lightyellow}1145.945 \\
36 & \cellcolor{lightyellow}\texttt{LIW:CONVERSATION} & \cellcolor{lightyellow}1350.528 & \cellcolor{lightorange}\texttt{ACT:TRENDS} & \cellcolor{lightorange}1172.241 & \cellcolor{lightpurple}\texttt{SEN:NEG-EMOJI} & \cellcolor{lightpurple}1145.679 \\
37 & \cellcolor{lightteal}\texttt{WRI:PRON} & \cellcolor{lightteal}1348.809 & \cellcolor{lightteal}\texttt{WRI:FLESCH-KINKAID} & \cellcolor{lightteal}1168.360 & \cellcolor{lightpurple}\texttt{SEN:CPD} & \cellcolor{lightpurple}1144.436 \\
38 & \cellcolor{lightteal}\texttt{WRI:SPELL\_ERRORS} & \cellcolor{lightteal}1348.611 & \cellcolor{lightteal}\texttt{WRI:CONG} & \cellcolor{lightteal}1168.104 & \cellcolor{lightpurple}\texttt{SEN:NEU-EMOJI} & \cellcolor{lightpurple}1144.349 \\
39 & \cellcolor{lightyellow}\texttt{LIW:AFFECT} & \cellcolor{lightyellow}1348.413 & \cellcolor{lightblue}\texttt{SOC:PURITY} & \cellcolor{lightblue}1163.298 & \cellcolor{lightpurple}\texttt{SEN:SENT-EMOJI} & \cellcolor{lightpurple}1143.934 \\
40 & \cellcolor{lightteal}\texttt{WRI:INTJ} & \cellcolor{lightteal}1346.151 & \cellcolor{lightpurple}\texttt{SEN:SENT-EMOJI} & \cellcolor{lightpurple}1157.149 & \cellcolor{lightteal}\texttt{WRI:ADV} & \cellcolor{lightteal}1142.218 \\
41 & \cellcolor{lightteal}\texttt{WRI:IRONY} & \cellcolor{lightteal}1345.676 & \cellcolor{lightblue}\texttt{SOC:AUTHORITY} & \cellcolor{lightblue}1157.075 & \cellcolor{lightteal}\texttt{WRI:CONG} & \cellcolor{lightteal}1140.672 \\
42 & \cellcolor{lightpink}\texttt{TOX:THREAT} & \cellcolor{lightpink}1339.511 & \cellcolor{lightteal}\texttt{WRI:PROPN} & \cellcolor{lightteal}1156.584 & \cellcolor{lightorange}\texttt{ACT:MENTIONS\_LINKS} & \cellcolor{lightorange}1139.670 \\
43 & \cellcolor{lightteal}\texttt{WRI:ADV} & \cellcolor{lightteal}1338.324 & \cellcolor{lightteal}\texttt{WRI:SPELL\_ERRORS} & \cellcolor{lightteal}1154.185 & \cellcolor{lightyellow}\texttt{LIW:MOTIVATION} & \cellcolor{lightyellow}1139.582 \\
44 & \cellcolor{lightpurple}\texttt{SEN:NEG} & \cellcolor{lightpurple}1337.666 & \cellcolor{lightteal}\texttt{WRI:IRONY} & \cellcolor{lightteal}1151.980 & \cellcolor{lightteal}\texttt{WRI:DET} & \cellcolor{lightteal}1138.912 \\
45 & \cellcolor{lightblue}\texttt{SOC:GENERAL} & \cellcolor{lightblue}1331.506 & \cellcolor{lightorange}\texttt{ACT:MENTIONS\_LINKS} & \cellcolor{lightorange}1150.571 & \cellcolor{lightpink}\texttt{TOX:INSULT} & \cellcolor{lightpink}1137.612 \\
46 & \cellcolor{lightteal}\texttt{WRI:VERB} & \cellcolor{lightteal}1318.154 & \cellcolor{lightteal}\texttt{WRI:SMOG} & \cellcolor{lightteal}1146.108 & \cellcolor{lightblue}\texttt{SOC:DEMOGRAPHIC} & \cellcolor{lightblue}1137.423 \\
47 & \cellcolor{lightteal}\texttt{WRI:PROPN} & \cellcolor{lightteal}1313.759 & \cellcolor{lightyellow}\texttt{LIW:DRIVES} & \cellcolor{lightyellow}1145.194 & \cellcolor{lightyellow}\texttt{LIW:AFFECT} & \cellcolor{lightyellow}1136.799 \\
48 & \cellcolor{lightorange}\texttt{ACT:COMMENTS} & \cellcolor{lightorange}1310.718 & \cellcolor{lightblue}\texttt{SOC:HARM} & \cellcolor{lightblue}1143.599 & \cellcolor{lightpink}\texttt{TOX:SEVERE-TOXICITY} & \cellcolor{lightpink}1136.324 \\
49 & \cellcolor{lightpurple}\texttt{SEN:NEU} & \cellcolor{lightpurple}1305.519 & \cellcolor{lightteal}\texttt{WRI:PRON} & \cellcolor{lightteal}1143.352 & \cellcolor{lightpurple}\texttt{SEN:POS-EMOJI} & \cellcolor{lightpurple}1134.120 \\
50 & \cellcolor{lightteal}\texttt{WRI:NOUNS} & \cellcolor{lightteal}1304.647 & \cellcolor{lightyellow}\texttt{LIW:TEMPORAL} & \cellcolor{lightyellow}1141.857 & \cellcolor{lightyellow}\texttt{LIW:CULTURE} & \cellcolor{lightyellow}1133.908 \\
51 & \cellcolor{lightteal}\texttt{WRI:STOPWORDS} & \cellcolor{lightteal}1300.541 & \cellcolor{lightteal}\texttt{WRI:AUX} & \cellcolor{lightteal}1141.625 & \cellcolor{lightpink}\texttt{TOX:OBSCENE} & \cellcolor{lightpink}1133.743 \\
52 & \cellcolor{lightorange}\texttt{ACT:SUBMISSIONS} & \cellcolor{lightorange}1292.198 & \cellcolor{lightblue}\texttt{SOC:INGROUP} & \cellcolor{lightblue}1136.889 & \cellcolor{lightpink}\texttt{TOX:TOXICITY} & \cellcolor{lightpink}1133.316 \\
53 & \cellcolor{lightblue}\texttt{SOC:PURITY} & \cellcolor{lightblue}1274.784 & \cellcolor{lightpurple}\texttt{SEN:POS} & \cellcolor{lightpurple}1136.698 & \cellcolor{lightyellow}\texttt{LIW:LIFESTYLE} & \cellcolor{lightyellow}1130.407 \\
54 & \cellcolor{lightorange}\texttt{ACT:TRENDS} & \cellcolor{lightorange}1269.451 & \cellcolor{lightblue}\texttt{SOC:FAIRNESS} & \cellcolor{lightblue}1134.151 & \cellcolor{lightpurple}\texttt{SEN:NEU} & \cellcolor{lightpurple}1128.792 \\
55 & \cellcolor{lightteal}\texttt{WRI:NOVELTY} & \cellcolor{lightteal}1269.434 & \cellcolor{lightteal}\texttt{WRI:INTJ} & \cellcolor{lightteal}1132.721 & \cellcolor{lightyellow}\texttt{LIW:TEMPORAL} & \cellcolor{lightyellow}1128.338 \\
56 & \cellcolor{lightteal}\texttt{WRI:ADPOSITIONS} & \cellcolor{lightteal}1264.317 & \cellcolor{lightteal}\texttt{WRI:VERB} & \cellcolor{lightteal}1121.263 & \cellcolor{lightteal}\texttt{WRI:IRONY} & \cellcolor{lightteal}1126.011 \\
57 & \cellcolor{lightteal}\texttt{WRI:DET} & \cellcolor{lightteal}1253.132 & \cellcolor{lightteal}\texttt{WRI:SCONJ} & \cellcolor{lightteal}1115.567 & \cellcolor{lightteal}\texttt{WRI:NOVELTY} & \cellcolor{lightteal}1124.884 \\
58 & \cellcolor{lightyellow}\texttt{LIW:SUMMARY} & \cellcolor{lightyellow}1250.382 & \cellcolor{lightyellow}\texttt{LIW:SUMMARY} & \cellcolor{lightyellow}1113.113 & \cellcolor{lightteal}\texttt{WRI:INTJ} & \cellcolor{lightteal}1124.002 \\
59 & \cellcolor{lightteal}\texttt{WRI:ARTICLES} & \cellcolor{lightteal}1238.723 & \cellcolor{lightpurple}\texttt{SEN:NEU} & \cellcolor{lightpurple}1094.396 & \cellcolor{lightteal}\texttt{WRI:SPELL\_ERRORS} & \cellcolor{lightteal}1120.098 \\
60 & \cellcolor{lightpink}\texttt{TOX:INSULT} & \cellcolor{lightpink}1232.596 & \cellcolor{lightteal}\texttt{WRI:ADPOSITIONS} & \cellcolor{lightteal}1085.154 & \cellcolor{lightyellow}\texttt{LIW:SPATIAL} & \cellcolor{lightyellow}1118.784 \\
61 & \cellcolor{lightyellow}\texttt{LIW:SOCIAL} & \cellcolor{lightyellow}1219.891 & \cellcolor{lightpink}\texttt{TOX:THREAT} & \cellcolor{lightpink}1060.968 & \cellcolor{lightteal}\texttt{WRI:STOPWORDS} & \cellcolor{lightteal}1117.319 \\
62 & \cellcolor{lightpink}\texttt{TOX:TOXICITY} & \cellcolor{lightpink}1214.411 & \cellcolor{lightpurple}\texttt{SEN:CPD} & \cellcolor{lightpurple}1049.432 & \cellcolor{lightteal}\texttt{WRI:VERB} & \cellcolor{lightteal}1111.365 \\
63 & \cellcolor{lightyellow}\texttt{LIW:BIO} & \cellcolor{lightyellow}1212.964 & \cellcolor{lightgray}\texttt{EMB:HIDDEN} & \cellcolor{lightgray}948.059 & \cellcolor{lightteal}\texttt{WRI:PROPN} & \cellcolor{lightteal}1110.323 \\
64 & \cellcolor{lightgreen}\texttt{REL:SUBR} & \cellcolor{lightgreen}1155.066 & \cellcolor{lightpink}\texttt{TOX:ID-ATTACK} & \cellcolor{lightpink}922.653 & \cellcolor{lightteal}\texttt{WRI:ADPOSITIONS} & \cellcolor{lightteal}1100.310 \\
65 & \cellcolor{lightgreen}\texttt{REL:AUTHORS} & \cellcolor{lightgreen}1142.419 & \cellcolor{lightpurple}\texttt{SEN:NEG} & \cellcolor{lightpurple}908.108 & \cellcolor{lightyellow}\texttt{LIW:CONVERSATION} & \cellcolor{lightyellow}1099.490 \\
66 & \cellcolor{lightpink}\texttt{TOX:ID-ATTACK} & \cellcolor{lightpink}1129.271 & \cellcolor{lightpink}\texttt{TOX:SEVERE-TOXICITY} & \cellcolor{lightpink}889.121 & \cellcolor{lightpurple}\texttt{SEN:POS} & \cellcolor{lightpurple}1095.805 \\
67 & \cellcolor{lightgreen}\texttt{REL:THREADS} & \cellcolor{lightgreen}1110.959 & \cellcolor{lightpink}\texttt{TOX:OBSCENE} & \cellcolor{lightpink}825.793 & \cellcolor{lightteal}\texttt{WRI:NOUNS} & \cellcolor{lightteal}1093.251 \\
68 & \cellcolor{lightgray}\texttt{EMB:HIDDEN} & \cellcolor{lightgray}1025.402 & \cellcolor{lightpink}\texttt{TOX:TOXICITY} & \cellcolor{lightpink}792.252 & \cellcolor{lightgreen}\texttt{REL:SUBR} & \cellcolor{lightgreen}1082.163 \\
69 & \cellcolor{lightgreen}\texttt{REL:ENGAGEMENT} & \cellcolor{lightgreen}959.841 & \cellcolor{lightpink}\texttt{TOX:INSULT} & \cellcolor{lightpink}787.978 & \cellcolor{lightgray}\texttt{EMB:HIDDEN} & \cellcolor{lightgray}986.320 \\
    \bottomrule
    \caption{Feature subgroup order of inspection for each task. Feature subgroups are color-coded according to their parent group (identified by the prefix before the colon).}
    \label{tab:featorder}
    \end{longtable}
    %}
\end{center}
\normalsize

%\begin{table}[ht!]
% \centering
% \resizebox{\textwidth}{!}{
% \input{tables/auc_featuorder_EMQ_5_classes}
% }
% \caption{Feature block order of inspection for each dataset. Feature blocks are colour-coded according to their feature group (identified by the prefix before the colon).}
% \label{tab:featorder}
%\end{table}

%\section{Feature Description}
%The following table presents the glossary of the various subgroups of features.
%\input{tab-glossary}

\end{document}